\documentclass[%
twocolumn,amsmath,amssymb,linenumber,
reprint,
superscriptaddress,
amsmath,amssymb,
prb,
]{revtex4-1}

\usepackage{graphicx}
\usepackage{dcolumn}
\usepackage{bm}
\usepackage{float}
\usepackage{threeparttable}
\usepackage{multirow}

\def\beq{\begin{equation}}
\def\eeq{\end{equation}}
\def\bea{\begin{eqnarray}}
\def\eea{\end{eqnarray}}
\def\brcl{\begin{array}{rcl}}
\def\bccl{\begin{array}{ccl}}
\def\blcl{\begin{array}{lcl}}
\def\err{\end{array}}
\def\fatR{{\bf R}}
\def\fatr{{\bf r}}

\begin{document}

\preprint{APS/123-QED}

\title{
Fast and accurate predictions of covalent bonds in chemical space
}
\author{K. Y. Samuel Chang}
\affiliation{Institute of Physical Chemistry and National Center for Computational Design and Discovery of Novel Materials (MARVEL), Department of Chemistry, University of Basel, 4056 Basel, Switzerland}
\author{Stijn Fias}
\affiliation{General Chemistry (ALGC), Free University Brussels (VUB), Pleinlaan 2, 1050 Brussel, Belgium}
\author{Raghunathan Ramakrishnan}
\affiliation{Institute of Physical Chemistry and National Center for Computational Design and Discovery of Novel Materials (MARVEL), Department of Chemistry, University of Basel, 4056 Basel, Switzerland}
\author{O. Anatole von Lilienfeld}
\email{anatole.vonlilienfeld@unibas.ch}
\affiliation{Institute of Physical Chemistry and National Center for Computational Design and Discovery of Novel Materials (MARVEL), Department of Chemistry, University of Basel, 4056 Basel, Switzerland}

\date{\today}

\begin{abstract}
We assess the predictive accuracy of perturbation theory based estimates of changes in covalent bonding 
due to linear alchemical interpolations among molecules. 
We have investigated $\sigma$ bonding to hydrogen, as well as $\sigma$ and $\pi$ bonding between main-group elements,
occurring in small sets of iso-valence-electronic molecular species with elements drawn from second to fourth rows in the $p$-block of the periodic table.
Numerical evidence suggests that first order estimates of covalent bonding potentials can achieve
chemical accuracy if (i) the alchemical interpolation is vertical (fixed geometry), 
(ii) involves molecules containing elements in the third and fourth row of the periodic table, 
and (iii) a reference geometry is optimized.
In this case, changes in the bonding potential become near-linear in coupling parameter,
resulting in analytical predictions with very high accuracy ($\sim$1 kcal/mol). 
Second order estimates deteriorate the prediction. 
If initial and final molecules differ not only in composition but also in geometry, 
all estimates become substantially worse,
with second order being slightly more accurate than first order.
The independent particle approximation to the second order perturbation 
performs poorly when compared to the coupled perturbed or finite difference approach.
Taylor series expansions up to fourth order of the potential energy curve 
of highly symmetric systems indicate a finite radius of convergence, 
as illustrated for the alchemical stretching of H$_2^+$. 
Numerical results are presented for (i) covalent bonds to hydrogen
in 12 molecules with 8 valence electrons (CH$_4$, NH$_3$, H$_2$O, HF, 
SiH$_4$, PH$_3$, H$_2$S, HCl,
GeH$_4$, AsH$_3$, H$_2$Se, HBr);
(ii) main-group single bonds in 9 molecules with 14 valence electrons
(CH$_3$F, CH$_3$Cl, CH$_3$Br,
SiH$_3$F, SiH$_3$Cl, SiH$_3$Br,
GeH$_3$F, GeH$_3$Cl, GeH$_3$Br);
(iii) main-group double bonds in 9 molecules with 12 valence electrons 
(CH$_2$O, CH$_2$S, CH$_2$Se,
SiH$_2$O, SiH$_2$S, SiH$_2$Se,
GeH$_2$O, GeH$_2$S, GeH$_2$Se);
(iv) main-group triple bonds in 9 molecules with 10 valence electrons
(HCN, HCP, HCAs,
HSiN, HSiP, HSiAs,
HGeN, HGeP, HGeAs); 
(v) H$_2^+$ single bond with 1 electron.
\end{abstract}

\pacs{Valid PACS appear here}
\maketitle


\section{Introduction}
Solving Schr\"odinger's time independent equation for the unperturbed electronic ground-state within the Born-Oppenheimer 
approximation yields the potential energy surface (PES) of {\em any} molecule
as a function of nuclear charges $\{Z_I\}$ (stoichiometry), 
nuclear positions $\{\fatR_I\}$ (geometry), and number of electrons $N$ (molecular charge).~\cite{Schrodinger_1926,Dirac_pcps_1929}
The PES plays a fundamental role in chemistry and elsewhere because many properties can be derived from it. 
While one can study efficient ways of predicting the PES of single compounds~\cite{HFfunctional_1985,HFfunctional_1989,dbdsmp}
efficient estimates of PES of ensembles of molecules are more useful (and challenging) in the context of virtual compound design
efforts.\cite{Beratan_science_1991,Zunger_1999,NorskovPRL2002,fpdesign2014anatole,RafaelMarques_jctc2015}
These efforts typically attempt to search chemical compound space (CCS) 
spanned by $\{\{Z_I\},\{\fatR_I\},N\}$\cite{Anatole_jcp_2006,Anatole_ijqc_2013} for novel materials with desirable properties. 
As such, accurate yet efficient quantum mechanics (QM) based PES estimates hold the key for successful rational compound design applications.\cite{Beratan_science_1991, Zunger_1999, Beratan_1996, Noble_science_2004, WYang_jcp_2007} 
While many inexpensive semi-empirical QM methods are available,
for this study we restrict ourselves to first principles in the spirit of 
Refs.~\cite{CarParrinello_prl_1985, Anatole_prl_2005, WYang_jacs_2006, Anatole_jcp_2007, Anatole_jcp_2009, Anatole_jcp_2010, Anatole_ijqc_2013, Reiher_ijqc_2014} 
More specifically, we investigate the application of ``alchemical'' coupling
to the problem of efficiently estimating the PES of new molecules 
using Taylor series expansions in CCS, rather than empiricism.

The alchemical coupling approach can be related to grand-canonical ensemble theory
(Widom insertion)~\cite{WidomInsertion1963,EMaginn-GCE_mp_1999, EMaginn-GCE_1999}, 
and has been well established for empirical force-field based molecular dynamics 
studies.~\cite{Jorgensen_jcp_1985, Gunsteren_jcamd_1987, Straatsma_arpc_1992, Gunsteren_pnas_2005,Oostenbrink_jpcb_2011} 
Using QM, alchemical changes are less common despite E.~B.~Wilson's early proposal of 
variable $Z$, back in 1962.~\cite{Wilson_jpc_1962}
Within QM, any two iso-electronic molecules in CCS can be coupled ``alchemically'' 
through interpolation of their external potentials.
Here, we have investigated if alchemical predictions can be used to model the 
PES of covalent bonds occurring in small closed-shell molecules made up from main group elements.
We have limited ourselves to covalent hydrogen bonds, as well as single, double, and triple bonds
in molecules with no more than 14 valence electrons.
We present and discuss numerical evidence for the following set of observations: 
First order Taylor-expansions of covalent bonding potentials can reach chemical
accuracy ($\sim$1 kcal/mol) if two conditions are met. 
Firstly, the alchemical change has to be ``vertical'', meaning
that initial reference molecule as well as final target molecule have to 
possess the same number of atoms located at the exact same positions. 
Secondly, all elements involved in the alchemical change, i.e.~all $\{Z_I\}$
destined to vary, have to occur late in the periodic table.
Second order Taylor-expansion based predictions are less accurate
than first order predictions if these conditions are met.
If reference and target molecule have different geometries, 
the predictive power of the first order Taylor expansion substantially deteriorates,
while second order estimates based on coupled perturbed Kohn-Sham equations
offer some improvement, however, without reaching chemical accuracy. 
Second order estimates based on the independent particle approximation
result in Taylor expansion estimates that are even worse than 
first order estimates.
For highly symmetrical alchemical changes, such as the dissociation of H$_2^+$, 
a finite radius of convergence is found.

In Sec.~\ref{sec:methods} we briefly summarize the framework of alchemical derivatives within Hartree-Fock and density
functional theory (DFT) as well as our notations. 
Numerical estimations of covalent bond stretching of small molecules are presented and discussed in Sec.~\ref{sec:results}: 
Extending previous work on alchemical perturbation,\cite{Anatole_jcp_2007,Anatole_jcp_2009,CHIMIA_2014} 
we discuss alchemical energy derivatives with respect to 
vertical transmutation, interpolating only the identity of the atoms while keeping the geometry fixed. 
Estimates of single, double and triple bonds are included as an application. 
We also report numerical results for alchemical stretching of chemical bonds using non-vertical transmutations.
Finally, conclusions are drawn in Sec.~\ref{sec:conclusions}. 

\section{Method}
\label{sec:methods}
\subsection{Taylor expansion in CCS}
A Taylor expansion in CCS can be constructed with the exclusive knowledge acquired by solving Schr\"odinger's equation for some reference molecule, with Hamiltonian $H_{\rm R}$, 
\beq\label{eq:TaylorExp}
E(\Delta \lambda) = E_{\rm R} + \Delta\lambda\partial_\lambda E_\lambda\Big\rvert_{\lambda=0} + \frac{\Delta\lambda^2}{2}\partial_\lambda^2 E_\lambda\Big\rvert_{\lambda=0} + \cdots.
\eeq

Derivatives of the total potential energy can be obtained by coupling a reference 
Hamiltonian to some target Hamiltonian, $H_{\rm T}$, 
such that $H_\lambda$ transforms $H_{\rm R}$ into $H_{\rm T}$ 
\beq\label{eq:linearH}
H_\lambda = (1-\lambda)H_{\rm R} + \lambda H_{\rm T},
\eeq
as the coupling parameter $\lambda$ goes from 0 to 1.
And consequently, 
$\partial_\lambda^m E_\lambda = \partial_\lambda^m\langle H_\lambda\rangle$, 
with $\partial_\lambda H_\lambda = H_{\rm T}-H_{\rm R} = H'$ being the alchemical perturbing Hamiltonian.
If these derivatives can be computed, $E_{\rm T}$ can be estimated according to Eq.~(\ref{eq:TaylorExp})
by setting $\Delta\lambda = 1$. 
Note that we couple reference and target systems in a linear and global fashion.
This is an arbitrary choice, non-linear and local interpolation functions could have
been chosen just as well.
In fact, in Ref.~\onlinecite{Anatole_jcp_2009}, an empirical quadratic interpolation function is found
to yield superior results for first order predictions of highest occupied molecular orbital (HOMO) eigenvalues.
In this study of alchemical changes of covalent bonding, 
we begin with linear and global interpolations, future work might deal with alternative
functions.

Given a pair of isoelectronic reference/target systems, described by $\{\{Z_I^{\rm R}\},\{\fatR_I^{\rm R}\},N\}$ and $\{\{Z_I^{\rm T}\},\{\fatR_I^{\rm T}\},N\}$ respectively, one can couple the two systems such that certain $Z_I^{\rm R}$ and $Z_I^{\rm T}$ are paired. 
Note that $Z_I^{\rm R}$ or $Z_I^{\rm T}$ can be scaled down to/up from zero if the number of atoms in one molecule is smaller.
Under isoelectronic conditions, the $\lambda$-dependent terms in 
the coupling Hamiltonian (Eq.~(\ref{eq:linearH}))
are the electron-nucleus and nucleus-nucleus interaction operators, 
\beq\label{eq:vext}
\bccl
v_\lambda(\fatr) &=&\displaystyle \sum_{I}^{N_I}\bigg(-\frac{(1-\lambda)Z_I^{\rm R}}{|\fatr-\fatR_I^{\rm R}|}-\frac{\lambda Z_I^{\rm T}}{|\fatr-\fatR_I^{\rm T}|}\bigg),
\\[12pt]
V_\lambda &=&\displaystyle \sum_{I< J}^{N_I}\bigg(\frac{(1-\lambda)Z_I^{\rm R}Z_J^{\rm R}}{|\fatR_I^{\rm R}-\fatR_J^{\rm R}|}+\frac{\lambda Z_I^{\rm T}Z_J^{\rm T}}{|\fatR_I^{\rm T}-\fatR_J^{\rm T}|}\bigg).
\err
\eeq
Since different pairing schemes result in different $v_\lambda(\fatr)$ and $V_\lambda$, 
it is obvious that the alchemical perturbation is alignment dependent. 
To investigate the behaviour of higher order corrections and the effects of varying geometry/stoichiometry, we neglect all relaxation effects for vertical iso-valence-electronic changes (see Sec.~\ref{sec:computation_detail}).

\subsection{First order derivative}
The first order derivative of the energy with respect to an alchemical interpolation parameter 
connecting {\em any} two iso-electronic molecules, 
can be computed according to the Hellmann-Feynman theorem,\cite{Feynman_pr_1939} 
as shown for molecular HOMO eigenvalues,\cite{Anatole_jcp_2009}
\beq\label{eq:d1E}
\partial_\lambda E_\lambda = \langle\partial_\lambda H_\lambda\rangle_\lambda
=\int d\fatr\: \rho_\lambda(\fatr) \partial_\lambda v_\lambda(\fatr) + \partial_\lambda V_\lambda,
\eeq
where $\rho_\lambda(\fatr)$ denotes the electron density, dependent on $\lambda$. 
At $\lambda=0$ we have $\rho_\lambda(\fatr)=\rho_{\rm R}(\fatr)$, 
which is independent of the target system. 
As such, the first order derivative can be calculated with a single reference density 
and without additional self-consistent field (SCF) calculation for {\em any} target system. 
In several circumstances, Taylor expansion estimates using first order alchemical derivatives 
have shown good accuracy for the rapid prediction of properties throughout CCS.\cite{Anatole_jctc_2007, Anatole_jcp_2010, Geerlings_jctc_2013, Anatole_ijqc_2013, CHIMIA_2014} 
In general, however, first order derivatives might not be sufficient.
Taking higher order derivatives into account might offer higher accuracy, 
assuming Eq.~(\ref{eq:TaylorExp}) converges rapidly. 

\subsection{Second order derivative}
Differentiation of Eq.~(\ref{eq:d1E}), based on linear interpolated Hamiltonian in Eq.~(\ref{eq:linearH}), yields
\beq\label{eq:d2E}
\partial_\lambda^2 E_\lambda = \int d\fatr\:\big(\partial_\lambda\rho(\fatr)\big)\big(\partial_\lambda v_\lambda(\fatr)\big),
\eeq
requiring the density response due to the alchemical perturbation.
Again, at $\lambda=0$ this amounts to the density response of the reference system. 
Evaluation of Eq.~(\ref{eq:d2E}) implies a differing density response for each target system. 
We have considered three approximations to $\partial_\lambda \rho$ including 
second order perturbation theory with independent particle approximation\cite{Kohn_prl_1985} (IPA), 
coupled perturbed (CP) approaches,\cite{WYang_jcp_2012, Geerlings_jctc_2010} 
as well as finite difference approximation (FD). 
Note that Eq.~(\ref{eq:d2E}) can be rewritten as
$\partial^2_\lambda E=\int d\fatr d\fatr' (\partial_\lambda v(\fatr))(\partial_\lambda v(\fatr'))\frac{\delta^2 E}{\delta v(\fatr)\delta v(\fatr')}$, 
where $\frac{\delta^2E}{\delta v(\fatr)\delta v(\fatr')}=\chi(\fatr,\fatr')$ 
is the static linear response function or susceptibility, well established within conceptual DFT~\cite{Parr_DFT,Sebastiani_2000,Galli_prb_2008,  Geerlings_jpcl_2010, Geerlings_jpca_2013, Geerlings_csr_2014}.

Perturbation theory provides ways to estimate $\partial_\lambda\rho_\lambda(\fatr)$.\cite{DFPT} 
Within IPA~\cite{Adler_1962, Wiser_1963, Kohn_prl_1985}, the static density response for a close-shell system is approximated by
\beq\label{eq:dn_rpa}
\brcl
\partial_\lambda\rho_\lambda(\fatr) &\approx&\displaystyle
-4\sum_{ia}\phi_{i}(\fatr)\phi_{a}(\fatr)
\\[8pt]&&\displaystyle\times
\int d\fatr'\:\frac{\phi_{i}(\fatr')\phi_{a}(\fatr')}{\varepsilon_{a} -\varepsilon_{i}}\partial_\lambda v_\lambda(\fatr'),
\err
\eeq
where $\{\phi_{i},\varepsilon_{i}\}$ denote the $i^{th}$ occupied molecular orbitals (MOs) and their eigenvalues, while $\{\phi_{a},\varepsilon_{a}\}$ denote the $a^{th}$ unoccupied counterparts. 
IPA neglects the influence of the alchemical perturbation on the 
Hartree and exchange-correlation (xc) potentials.\cite{Geerlings_jctc_2010, Geerlings_jpcl_2010} 
Note that Eq.~(\ref{eq:dn_rpa}) becomes numerically exact for 1-electron system with converged 
basis set within Hartree-Fock approximation, because of the absence of Coulomb and xc interaction between electrons.

Recently, Yang, Cohen, De Proft and Geerlings derived an expression of the density response that also includes the dependence of Coulomb and xc potential,\cite{WYang_jcp_2012} the CP approach,\cite{Geerlings_csr_2014}
\beq\label{eq:dn_cpks}
\brcl
\partial_\lambda\rho_\lambda(\fatr) &=& \displaystyle
-4\sum_{ij}\sum_{ab}\phi_{i}(\fatr)\phi_{a}(\fatr)
\\[8pt]&&\displaystyle\times
(\mathbf{M}^{-1})_{ia,jb}\int d\fatr'\:\phi_{j}(\fatr')\phi_{b}(\fatr')\partial_\lambda v_\lambda(\fatr'),
\err
\eeq
where the matrix elements of $\mathbf{M}$ are
\beq\left\lbrace\label{eq:dn_cpks_matrixElements}
\blcl
\mathbf{M}_{ia,jb} &=& (\varepsilon_{a}-\varepsilon_{i})\delta_{ij}\delta_{ab}
+4\mathbf{J}_{ia,jb} +4 \mathbf{X}_{ia,jb},
\\[8pt]
\mathbf{J}_{ia,jb}&=&\displaystyle\int d\fatr d\fatr'\frac{\phi_{i}(\fatr)\phi_{a}(\fatr)\phi_{j}(\fatr')\phi_{b}(\fatr')}{|\fatr-\fatr'|},
\\
\mathbf{X}_{ia,jb}&=&\displaystyle \int d\fatr d\fatr'\phi_{i}(\fatr)\phi_{a}(\fatr)\phi_{j}(\fatr')\phi_{b}(\fatr')
\\[8pt]&&\displaystyle\times
\bigg(\frac{\delta^2 E_{xc}}{\delta \rho(\fatr)\delta \rho(\fatr')}\bigg).
\err\right.
\eeq
In the limit of $\mathbf{J}_{ia,jb}\rightarrow 0$ and $\mathbf{X}_{ia,jb}\rightarrow 0$, Eq.~(\ref{eq:dn_rpa}) and Eq.~(\ref{eq:dn_cpks}) are equivalent.

Alternatively, one can also introduce an explicit small perturbation and converge the new density at ${\Delta\lambda \ll 1}$. The density response can then be estimated via FD, ${\partial_\lambda\rho(\fatr)\approx \frac{\rho_{\Delta\lambda}(\fatr)-\rho_{\rm R}(\fatr)}{\Delta\lambda}}$. 
In practice, instead of starting the SCF for the perturbed system from atom based initial guesses, 
we restart with $\rho_{\rm R}(\fatr)$ resulting in convergence within few SCF steps. 

\subsection{Higher order derivatives}
M{\o}ller-Plesset (MP) perturbation theory\cite{MP2, Cremer_advRev_2012} 
is used to estimate correlation energy corrections based on converged Hartree-Fock results. 
The derivation of higher order corrections in MP theory are equivalent to the $m^{th}$ order alchemical derivative. 
Here, instead of the two-particle operator for electron-electron interaction as perturbation in MP theory, 
the alchemical perturbation operator $H_T - H_R$ can be used.
Within IPA, the MP formula can be directly applied to obtain any $m^{th}$ order derivative.

\subsection{Predicting changes in covalent bonds}
For the study of covalent bonds we focus on the 
changes in binding potential due to alchemical coupling.
We consider the difference in total potential energy between 
two bounded atoms at two arbitrary interatomic distances $d$ and $d_0$,
\beq
\label{eq:revE}
\Delta E(d,d_0) =  E(d) - E(d_0).
\eeq
If, for example, $d_0$ is large and $d$ is the geometry minimum,
$\Delta E$ becomes the bond dissociation energy.
We are interested in changes of $\Delta E(d,d_0)$ as a function of $d$ 
due to alchemical changes for a fixed $d_0$. 
More specifically, we couple a reference to target system
via the corresponding Hamiltonians yielding expectation values as a function
of $\lambda$, 
\bea
\label{eq:revE_lambda}
\Delta E_\lambda(d,d_0) &= & E_\lambda(d)  - E_\lambda(d_0) \\
&= & \langle H_R(d) + \lambda (H_T(d)-H_R(d)) \rangle \nonumber\\
&&- \langle H_R(d_0) + \lambda (H_T(d_0)-H_R(d_0)) \rangle. \nonumber
\eea
As $\lambda$ goes from zero to one, the two components in Eq.~(\ref{eq:revE}) change 
from reference ($E_{\rm R}(d), E_{\rm R}(d_0)$) 
to target ($E_{\rm T}(d), E_{\rm T}(d_0)$) compound.
The truncated Taylor expansion based estimate of the target compound's
potential is then obtained via, 
\bea
\label{eq:Model}
\Delta E_{\rm T}(d,d_0) \approx \Delta E_{\rm T}^{(m)}(d,d_0) & = & \Delta E_{\rm R}(d,d_0)\\
&&+ \sum_{k=1}^m \frac{1}{k!}\partial_\lambda^k \Delta E_\lambda(d,d_0), \nonumber 
\eea 
where the superscript $m$ stands for Taylor expansion with $m$ terms,
as a function of bond-length $d$ for vertical alchemical changes. Since $\Delta E_{\rm T}$ is the property of interest, the subscript T, $\lambda$, and the dependency of $d_0$ will be omitted for the rest of this work, unless otherwise noted. 
In this study we investigated orders up to $m=4$ for the stretching of H$_2^+$, 
and up to $m=2$ for all other molecules.
For a fixed $d_0$, first and second order estimation are
\bea
\label{eq:delta_E_explicit_1st}
\Delta E^{(1)}(d) &=& \big(E_{\rm R}(d) + \partial_\lambda E_{\rm \lambda}(d)\big) 
\\&&- \big(E_{\rm R}(d_0) + \partial_\lambda E_{\rm \lambda}(d_0)\big),
\nonumber\\
\label{eq:delta_E_explicit_2nd}
\Delta E^{(2)}(d) &=&\big(E_{\rm R}(d) + \partial_\lambda E_{\rm \lambda}(d) + \frac{1}{2}\partial^2_\lambda E_{\rm \lambda}(d)\big) 
\\&&- \big(E_{\rm R}(d_0) + \partial_\lambda E_{\rm \lambda}(d_0)+\frac{1}{2}\partial^2_\lambda E_{\rm \lambda}(d_0)\big).
\nonumber
\eea
Since $d$ and $d_0$ in Eq.~(\ref{eq:revE}) are arbitrary, 
one can infer the binding curve via scanning $d$ for any fixed $d_0$. 
The predictive power, however, happens to dependent on $d_0$. 
For this reason, we optimize $d_0$ such that the integrated error in dissociation region is minimal. As shown in Fig.~\ref{fig:d_scatter}, an empirical linear relationship exists between equilibrium bond length of target molecule $d_{\rm eq}^{\rm T}$, and $d_{\rm opt}$
\beq\label{eq:d_opt}
d_{\rm opt}\approx 0.76\: d_{\rm eq}^{\rm T} + 0.97\:{\rm \AA}.
\eeq
$d_0$ is determined according to Eq.~(\ref{eq:d_opt}) for all vertical changes. If $d_{\rm eq}^{\rm T}$ is not known it can easily be estimated with semi-empirical quantum chemistry methods.
For non-vertical changes, 
we fix $d_0 = d_{\rm eq}$ to the equilibrium distance in the reference molecule, resulting in $\Delta E^{(m)}(d_{\rm eq})=0$.
Eqs.~(\ref{eq:delta_E_explicit_1st}, \ref{eq:delta_E_explicit_2nd}) become
\bea
\Delta E^{(1)}(d) &=& \big(E_{\rm R}(d) + \partial_\lambda E_{\rm \lambda}(d)\big) 
\\&&- \big(E_{\rm R}(d_{\rm eq}) + \partial_\lambda E_{\rm \lambda}(d_{\rm eq})\big),
\nonumber\\
\Delta E^{(2)}(d) &=&\big(E_{\rm R}(d) + \partial_\lambda E_{\rm \lambda}(d) + \frac{1}{2}\partial^2_\lambda E_{\rm \lambda}(d)\big) 
\nonumber
\\&&- \big(E_{\rm R}(d_{\rm eq}) + \partial_\lambda E_{\rm \lambda}(d_{\rm eq})+\frac{1}{2}\partial^2_\lambda E_{\rm \lambda}(d_{\rm eq})\big).
\nonumber
\eea

\subsection{Error measures}
\label{sec:Error}
For bond lengths, we quantify the predictive power of the Taylor expansions by evaluating the deviation of prediction from the DFT bond length $\Delta d_{\rm eq} = d_{\rm eq}^{(m)}-d_{\rm eq}$, where $d_{\rm eq}^{(m)}$ stands for the predicted equilibrium distance of $\Delta E^{(m)}$. 
We calculate the deviation of the predicted energy at $d_{\rm eq}^{(m)}$ from the DFT energy at the DFT minimum, $\Delta E_{\rm eq} = \Delta E^{(m)}(d_{\rm eq}^{(m)}) - \Delta E(d_{\rm eq})$. 
The deviation in harmonic vibration frequency, $\Delta\omega = \omega^{(m)} - \omega$, 
of the stretching bond is also included in order to quantify the accuracy of the stiffness of 
the predicted binding potential. 
The vibration frequency is computed from the curvature of cubic spline interpolated binding potential,
$\omega = \frac{1}{2\pi}\sqrt{\frac{k_{\rm eq}}{\mu}}$ where $k_{\rm eq}=\partial_d^2 \Delta E(d_{\rm eq})$ and $\mu$ is the reduced mass.
%
%
Finally, we measure the integrated error (IE) for the dissociative tail, defined as
\beq
\mbox{IE} = \frac{1}{|d_{\rm max}-d_{\rm eq}^{(m)}|}\int_{d_{\rm eq}^{(m)}}^{d_{\rm max}}dx
|\Delta E^{(m)}(x)-\Delta E(x)|,
\eeq
for vertical iso-valence-electronic changes. 
Note that while in principle one would like $d_{\rm max} \rightarrow \infty$,
$d_{\rm max}$ has been set to correspond roughly to the inflection point, 
due to the issues of a single reference such as DFT method for describing covalent bond-dissociation. 
This shortcoming is also evident from comparison of DFT to CCSD(T) curves shown in Fig.~\ref{fig:pbe0vpbe}.

Note that this aspect is irrelevant for the alchemical predictions: 
If a more reliable reference method had been used the error integration could
easily be expanded to include the entire dissociative tail.
These four quantities provide a numerical indication of how good a prediction is. 
For a perfect prediction one would expect $(\Delta E_{\rm eq},\Delta d_{\rm eq}, \Delta\omega, \mbox{IE})=(0,0,0,0)$. 
Note that we compare the predictions to DFT. This is an arbitrary choice, any other QM method could 
have been applied just as well.  

\subsection{Computational details}
\label{sec:computation_detail}
Alchemical interpolations of molecules containing elements from different rows in the periodic table
can still be iso-electronic if effective core or pseudo potentials (PPs) are used, resulting
in a constant number of valence electrons~\cite{Anatole_ijqc_2013}.
For example, one can couple carbon to silicon using just four valence electrons. 
Non-local PPs are widely used to mimic the presence of core electrons in atoms~\cite{Kleinman_prl_1982},
and are amenable to the tuning of a wide range of properties
including dispersion forces, band-gap, or vibrational frequencies~\cite{Anatole_prl_2004,Anatole_prb_2008,FCACP_anatoleMP2013}.
The non-local external potential $v_\lambda(\fatr)$ in Eq.~(\ref{eq:vext}) then becomes
\beq\label{eq:vext_PPs}
v_\lambda(\fatr,\fatr')=\sum_{I}^{N_I}\Big((1-\lambda)v_{I}^{\rm R}(\fatr,\fatr')+\lambda v_I^{\rm T}(\fatr,\fatr')\Big),
\eeq
where $v_I^{\rm R}$ and $v_I^{\rm T}$ are PPs for $Z_I^{\rm R}$ and $Z_I^{\rm T}$ respectively. Note that $v_\lambda(\fatr,\fatr')$ in Eq.~(\ref{eq:vext_PPs}) and $v_\lambda(\fatr)$ in Eq.~(\ref{eq:vext}) result in different coupling Hamiltonians, and therefore different $\lambda$-dependencies of the energy and its derivatives. 

All results have been obtained within the Born-Oppenheimer approximation, 
where nuclei are clamped, 
nuclear repulsion $V_\lambda$ is decoupled from the electronic wavefunction, 
and is added as a geometry- and $\lambda$-dependent constant to the electronic energy.
Nuclear-nuclear repulsion energy is computed automatically by most QM codes. However, it must be removed and recomputed independently for $V_\lambda$ according to Eq.~(\ref{eq:vext}) to avoid self-repulsion between transmutating atoms. 
Throughout the present study, standard atomic and plane-wave basis functions, 
linearly interpolated PPs, as well as the PBE xc potential~\cite{PBE} within KS-DFT is used. 
The scanning of $0.5\:\mbox{\AA}\leq d\leq 3.0\:\mbox{\AA}$ 
is carried out with increments $\Delta d=0.1\:${\AA}.
For each prediction order $m$, $\Delta E^{(m)}(d)$ are interpolated with cubic splines, from which the stiffness $\partial_d^2\Delta E^{(m)}(d_{\rm eq}^{(m)})=k_{\rm eq}$ is computed.
All density volumetric data is printed into {\tt Gaussian} CUBE files, 
from which integrated density slices are calculated.

\subsubsection{Details for vertical iso-valence-electronic changes}
\label{sec:method_vertical}
Numerical results for vertical iso-valence-electronic alchemical changes 
(discussed in Sec.~\ref{sec:vertical} and \ref{sec:heavy}) 
have been obtained with {\tt CPMD}\cite{CPMD},
a plane wave basis with 100 Ry cutoff, and Goedecker PPs.\cite{Goedecker_1996, Goedecker_1998, Krack_2005} 
The periodic supercell size is 20$\times$15$\times$15\:\AA$^3$, 
and one heavy atom is fixed at (7.5\:\AA, 7.5\:\AA, 7.5\:\AA) 
while the stretching atom shifts along +$x$-axis. 
For each geometry, heavy atoms are mutated to other elements in the same column 
of the periodic table while all H are fixed at the same location as in the reference compound. 

Since Eq.~(\ref{eq:vext_PPs}) is a non-local operator, Eqs.~(\ref{eq:d1E}) and (\ref{eq:d2E}) 
need to be converted to wavefunction expressions. 
The first order derivative for the Hamiltonian $H_{{\rm R}\rightarrow{\rm T}}$ 
is evaluated using RESTART files in which the reference compound's density and wavefunctions 
have been stored: 
$\partial_\lambda E = \langle \partial_\lambda H\rangle_{\rm R} = E_{\rm T}[\rho_{\rm R}]-E_{\rm R}[\rho_{\rm R}]$. 
And the second order derivative is evaluated correspondingly relying on FD, 
$\partial_\lambda^2 E\approx \frac{\langle \partial_\lambda H\rangle_{\Delta\lambda}-\langle \partial_\lambda H\rangle_{\rm R}}{\Delta\lambda}$, with $\Delta\lambda=0.05$. Wavefunctions of reference compound are used for $\Delta E^{(1)}$, while $\Delta E^{(2)}_{\rm FD}$ is evaluated by FD with linearly interpolated PPs parameter. 

Coupled-cluster results (CCSD(T)) obtained for HCl and HBr 
have been computed using {\tt Gaussian09}\cite{Gaussian09} in aug-cc-pVTZ-\cite{aug-cc-pVTZ_H} basis, and default input parameters.

\subsubsection{Details for non-vertical iso-electronic changes}
Numerical results for non-vertical iso-electronic alchemical changes 
have been obtained using atom centered basis-sets.
Restricted open-shell Hartree-Fock calculations have been carried out using Cartesian aug-cc-pVTZ basis set\cite{aug-cc-pVTZ_H} for H$_2^+$ (Discussed in Sec.~\ref{sec:h2p}). Eq.~(\ref{eq:dn_rpa}) and higher order derivatives are evaluated analytically by Gaussian expansion of MOs. Reference geometry is first relaxed by {\tt Gaussian09}\cite{Gaussian09} and the converged MO coefficients are extracted to evaluate orbital integrals. {\tt NWChem}\cite{NWChem} is used to scan $\Delta E$ as a function of $\lambda$ in Fig.~\ref{fig:h2p}(d) along alchemical path with discretization $\Delta\lambda=0.01$. It is done by reassigning nuclear charges in the system. 

Non-vertical alchemical changes in 10-electron molecules (discussed in Sec.~\ref{sec:non-vertical}) 
have been calculated using the uncontracted Cartesian Def2TZVP basis set\cite{def2-tzvp}.
Uncontracted neon basis is used for second row heavy atoms. 
Additional hydrogen basis functions are placed along the stretching pathway, 
from $d=0.5$\:\AA\ to $d=3.0$\:\AA\ in increments $\Delta d=0.1$\:\AA. 
All systems with integer nuclear charges have been calculated using {\tt Gaussian09}\cite{Gaussian09} 
while systems with fractional nuclear charges have been calculated using {\tt NWChem}\cite{NWChem} 
with discretization $\Delta\lambda=0.01$. 
For each $0\leq\lambda\leq 1$, the atomic density for SCF initial guess iterates 
through \{C, N, O, F, Ne\} to ensure convergence. 
In all {\tt Gaussian} and {\tt NWChem} calculations we used Cartesian/Real spherical harmonic basis functions.

\begin{figure}[H]
\centering
\includegraphics[scale=0.4, angle=0, width=8.5cm]{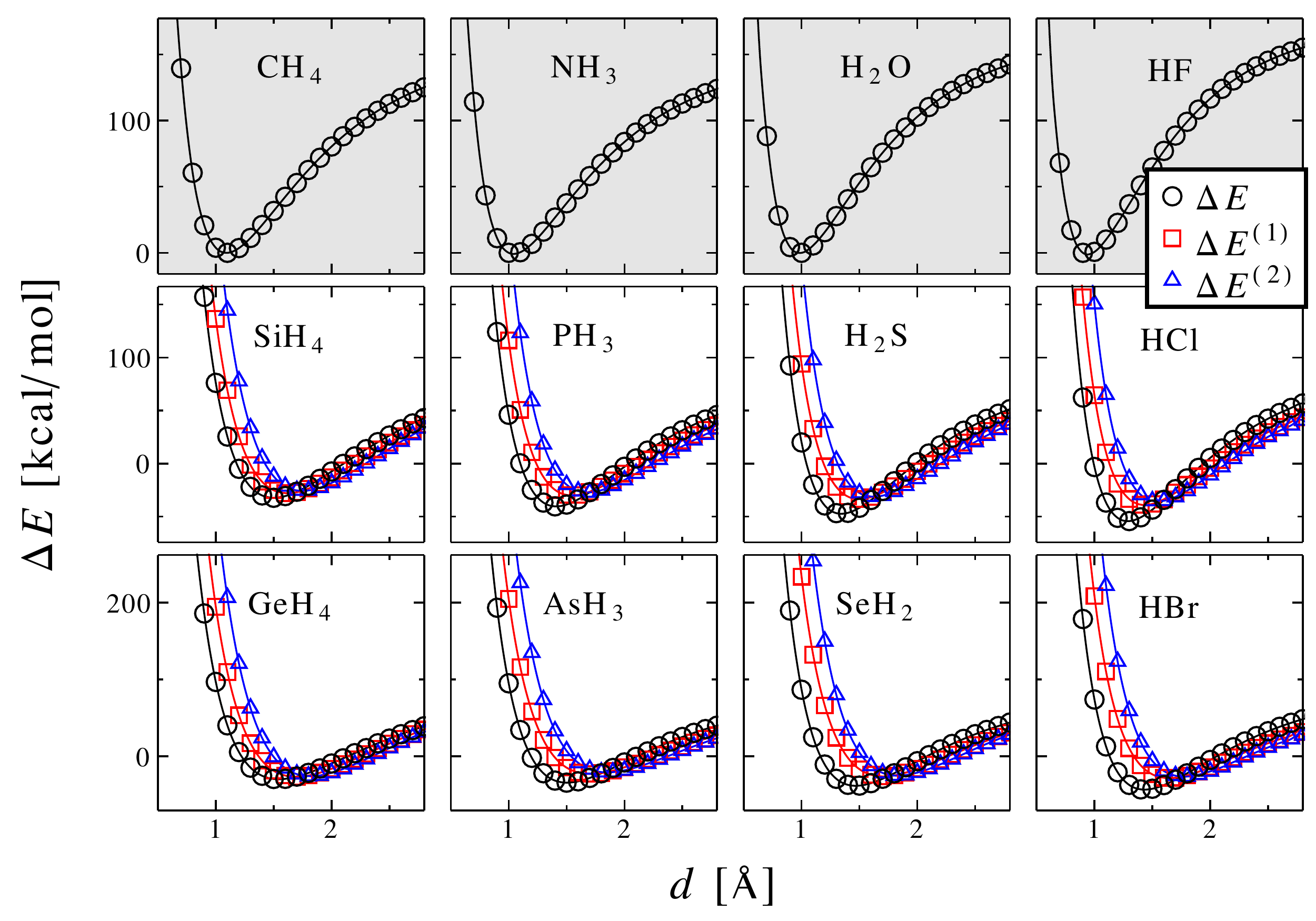}
\includegraphics[scale=0.4, angle=0, width=8.5cm]{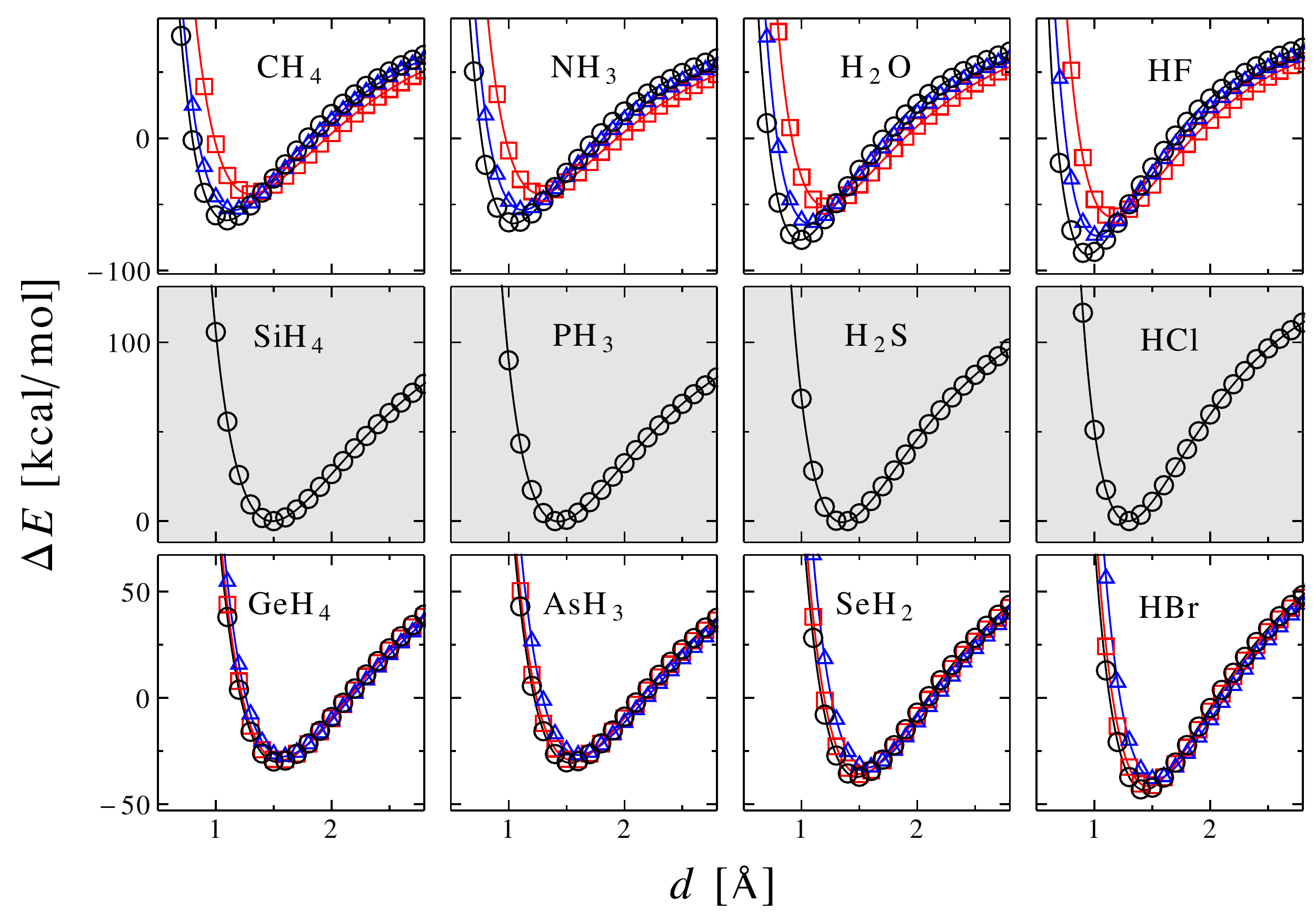}
\includegraphics[scale=0.4, angle=0, width=8.5cm]{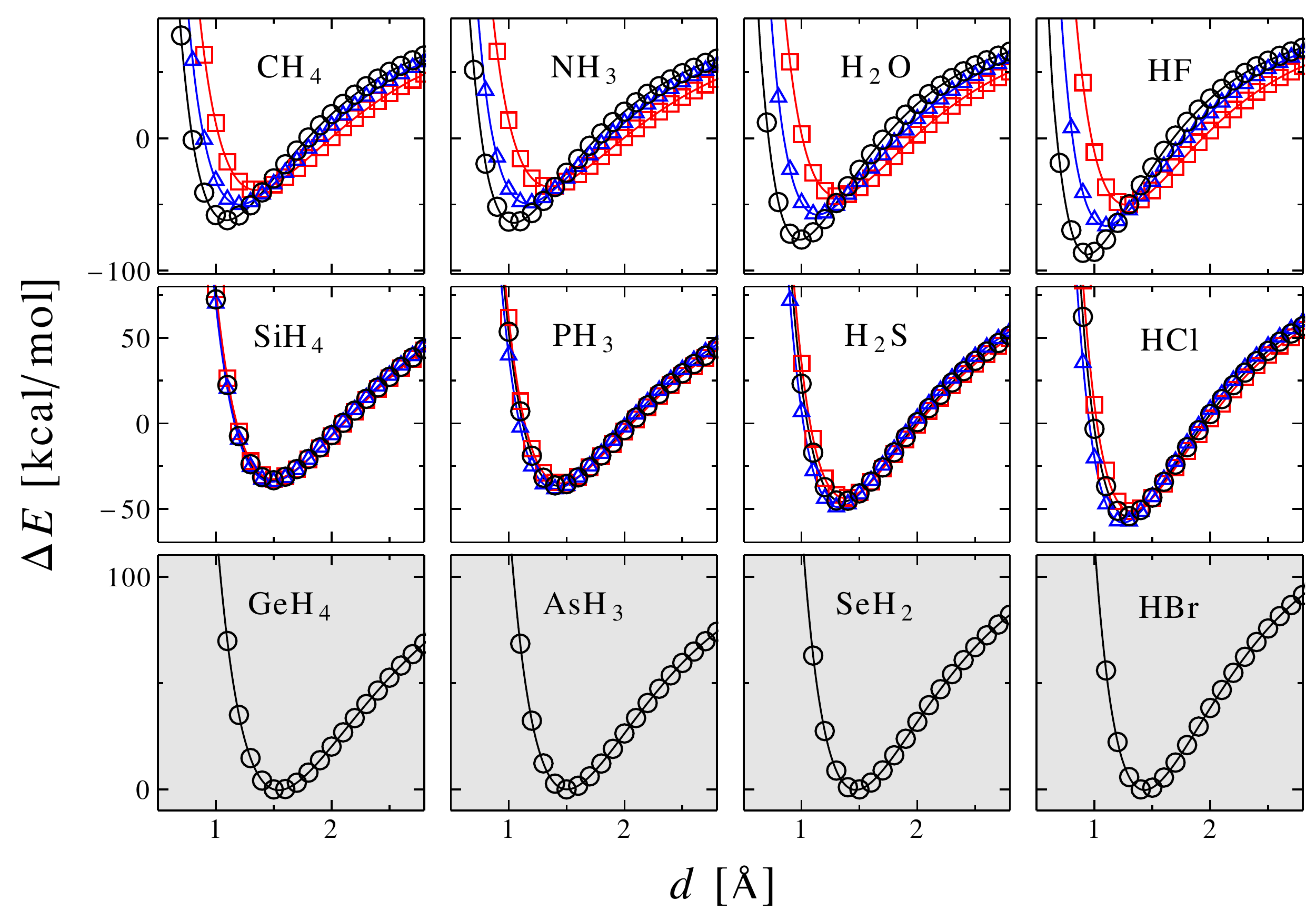}
\caption{
$\Delta E$ is shown as a function of $d$ in Eq.~(\ref{eq:Model}). 
White background panels: True (black circles), first (red squares) and second (blue triangles) order predictions 
of changes in the covalent bond of hydrogen due to vertical alchemical interpolations. 
Gray background panels: The true potentials of the reference compounds employed 
for the predictions for the first and second order predictions. 
}
\label{fig:vertical_H}
\end{figure}
\section{Results and Discussions}
\label{sec:results}
\subsection{Vertical iso-valence-electronic changes of X-H}
\label{sec:vertical}
\subsubsection{Predicted potentials}

Using Taylor expansions binding potentials have been estimated for
covalent bonds involving hydrogen (X-H) for the following 12 molecules with 8 valence electrons:
CH$_4$, NH$_3$, H$_2$O, HF (second period);
SiH$_4$, PH$_3$, H$_2$S, HCl (third period);
and GeH$_4$, AsH$_3$, H$_2$Se, HBr (fourth period).
Numerical results for vertical first (red) and second (blue) order truncated Taylor series estimates
feature in Fig.~\ref{fig:vertical_H}.
They measure the change in X-H binding energy as one goes from reference to target compound.

We first note that the entire potential is reproduced in semi-quantitative fashion 
for all combinations of reference/target molecules.
The precise predictive power strongly depends on the choice of reference/target molecule pair, 
on the choice of $d_0$, and on the expansion going up to first or second order.
First order estimates among molecules with elements from third or fourth row are 
very accurate (See Fig.~\ref{fig:vertical_H}, bottom and mid row in mid and bottom panel, respectively).
By contrast, predicting, or starting with, second row elements consistently yields worse results.
Inclusion of second order corrections does not necessarily lead to improved performance. 
Second order truncated Taylor series estimates only yield more accurate predictions than first order
when the reference molecule contains heavier elements than the target molecule.
For example, if we predict HF using HBr as a reference, the second order prediction is more accurate than first order.
Making the inverse prediction (i.e.~HBr from HF), however, 
first order is more accurate than second order.

\begin{figure}
\centering
\includegraphics[scale=0.4, angle=0, width=8.5cm]{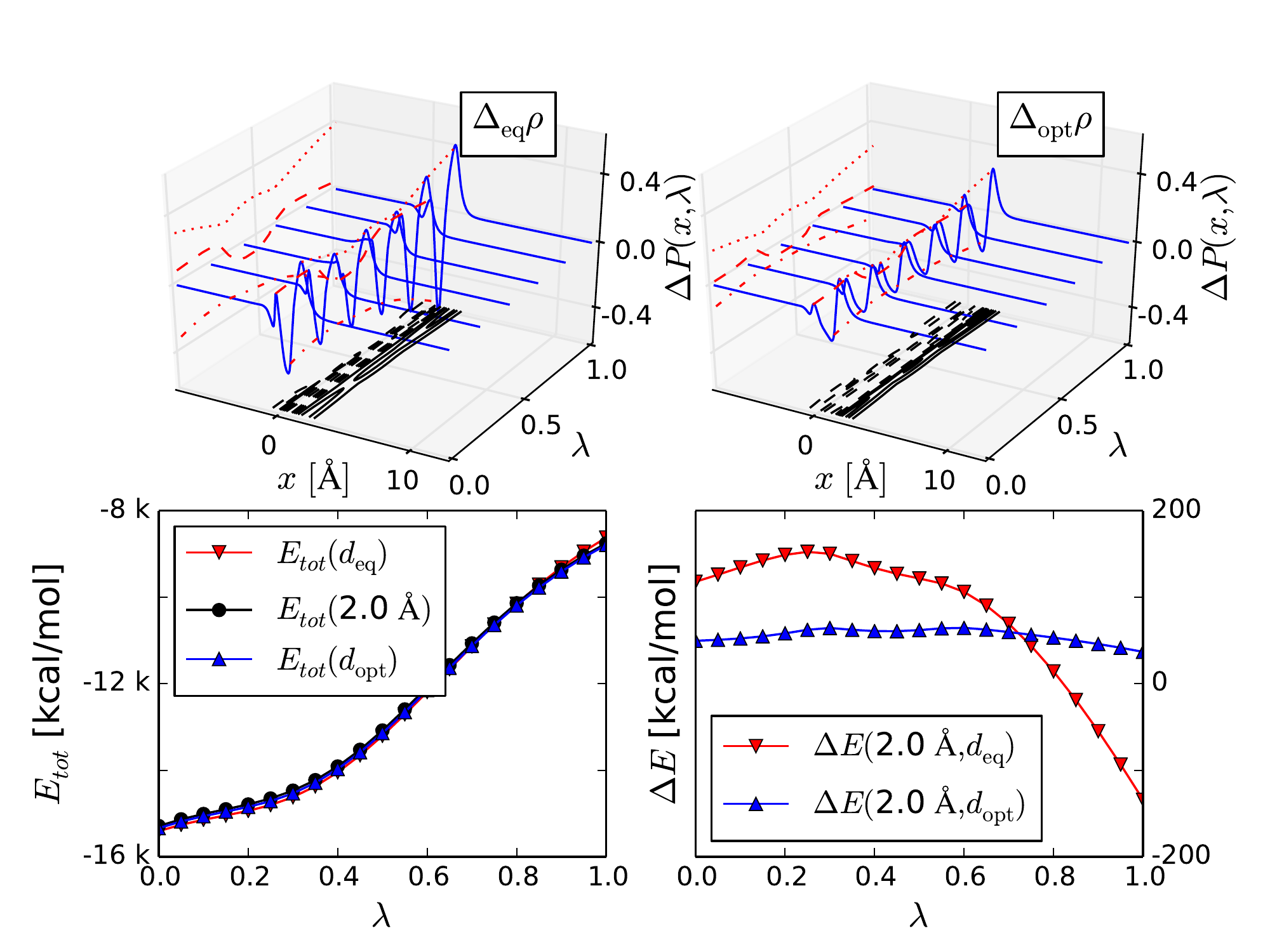}
\caption{
Alchemical coupling of HF ($\lambda=0$) to HBr ($\lambda = 1$).
TOP panel: 
$E$ and $\Delta E$ where $d_{\rm eq} = 0.94\:${\AA} (red)
denotes the equilibrium bond length of reference molecule HF, 
and $d_{\rm opt} = 1.57\:${\AA} (blue) linearizes $\Delta E$.
BOTTOM panel: 
Integrated valence electron density difference slices between H-X at $d = 2{\mbox{\AA}}$
and at $d_{\rm opt}$ and $d_{\rm eq}$, respectively,
$\Delta P_\lambda(x)=\int dy dz[\rho_\lambda(\fatr, d) - \rho_\lambda(\fatr, d_0)]$. 
Dependence on $\lambda$ is shown for the same vertical interpolations,
Left and Right corresponding to the non-linear (red) and linearized (blue) $\Delta E$ curves in right-hand TOP panel. 
Density changes at heavy atom and hydrogen positions are highlighted as red dashed and dotted/dash-dotted lines, respectively. 
}
\label{fig:p13d20}
\end{figure}

The performance of truncated Taylor series dramatically varies depending on the choice of the $d_0$ value.
The top panel in Fig.~\ref{fig:p13d20} illustrates this for $\Delta E(2\mbox{\AA}, d_0)$ 
for HF$\rightarrow$HBr as a function of $\lambda$, once with $d_0=0.94${\AA}---the equilibrium bond length of HF---
and once with $d_0=1.57${\AA}, a value for $d_0$ which happens to linearize $\Delta E$ in $\lambda$.
While the coupling path of total energies is hardly distinguishable for 
$E(2\mbox{\AA})$, $E (1.57\mbox{\AA})$, and $E (0.94\mbox{\AA})$, 
$\Delta E$ is strongly dependent on the choice of $d_0$.
By choosing $d_0=1.57\mbox{\AA}$, $\Delta E(2\mbox{\AA},1.57\mbox{\AA})$ in Eq.~(\ref{eq:revE_lambda}) 
becomes nearly linear, 
while plotting $\Delta E(2\mbox{\AA},0.94\mbox{\AA})$ reveals substantial curbing.
This is why, when choosing the right $d_0$, first order predictions of $\Delta E$ can be very predictive. 

The top panel in Fig.~\ref{fig:p13d20} also explains why second order estimates can be worse than first order, 
and why this changes for the reverse coupling:
On the side of the lighter element ($\lambda \rightarrow$ 0), 
a weak convexity is noticable in $\Delta E$ (blue line), despite the overall concavity of the path. 
The presence of inflection points will always lead to a deterioration of second order predictions,
resulting in a more accurate first order estimate. 
On the other side ($\lambda \rightarrow$ 1), no such inflection point exists and the second order term
results in the expected improvement of the prediction.
For the other compound pairs shown in Fig.~\ref{fig:vertical_H} similar observations can be 
made for the attractive part of the bonding potential 
(see also $\Delta E(2\mbox{\AA},d_0)$ curves in supplementary materials).
We have found that the inflection point near $\lambda=0$ occurs 
only when the reference molecule has the lighter element.
Conversely, no inflection point has been observed for atoms transmutating upward the column. 
We believe that this behaviour is due to the specifics of the employed PPs. 
Future studies will show why this is the case, and if similar trends hold for other PPs.

\subsubsection{Integrated error}
Prediction errors for energy minima, equilibrium bond lengths, force constants, and integrated error in dissociation region (calculated as described in Sec.~\ref{sec:Error})
have been obtained for all predictions in Fig.~\ref{fig:vertical_H},
and are listed in Table.~\ref{table:vertical_error}. 
The results lend quantitative support for the observations articulated above. 
In particular, the results suggest that chemical accuracy can be obtained when using 
first order Taylor series based estimates among compounds containing third and fourth row elements.
Second order based predictions are {\em always} worse except when a molecule with heavier 
element is used as a reference to predict a molecule with lighter one, for example HBr$\rightarrow$HF. 

The best prediction performance is found for first order based estimates 
using reference molecules containing third row elements ($n_{\rm R}=3$) 
in order to predict target molecules made up of fourth row elements ($n_{\rm T}=4$).
The overall average deviation from reference bonding potential energies and integrated error are $\sim$2.5 kcal/mol.
Corresponding predictions of equilibrium distances deviate at most by 0.03{\AA}, 
and the vibration frequencies deviate no more than 32 cm$^{-1}$.
Second order estimates for the same third and fourth row combinations give slightly worse results.
The worst predictions are found if the coupled molecules skip a row, 
{\it i.e.} involve elements from second and fourth row---for 
first as well as second order truncated Taylor expansions. 
This is not surprising as the central atom's electron density must accommodate the most severe 
contractions/expansions for such interpolations. 
Moreover, second order overcorrections can also be found (Table.~\ref{table:vertical_error})
whenever molecules containing fourth row elements are used to predict molecules containing third row elements: 
Both, predicted energy minimum and equilibrium bond length, show negative deviations.

\begin{widetext}

\begin{table}[H]
\def\thsp{0.6pc}
\caption{
Error measures for first (left) and second (right) order predictions of vertical iso-valence-electronic alchemical changes 
of covalent bond potentials in X-H$\rightarrow$A-H. 
The compound pairs are arranged in the same way as in Fig.~\ref{fig:vertical_H}, 
i.e.~each box corresponds to an unshaded panel containing a predicted potential.
Unit are [kcal/mol] for $\Delta E_{\rm eq}$ and IE, 
[\AA] for $\Delta d_{\rm eq}$, and $\Delta\omega$ in [cm$^{-1}$].
Avg corresponds to averaged signed error for each row with corresponding unit.
Periods of X and A are denoted by their respective primary quantum number $n_{\rm R}, n_{\rm T}=\{2,3,4\}$.
IV, V, VI, VII represent the columns of X and A in the periodic table. 
}
\begin{threeparttable}
\begin{minipage}{\textwidth}
\centering
\begin{minipage}{0.45\textwidth}
\begin{tabular}{l>{\hspace{\thsp}}l|
>{\hspace{\thsp}}r|
>{\hspace{\thsp}}r|
>{\hspace{\thsp}}r|
>{\hspace{0.3pc}}r||
>{\hspace{0.2pc}}r}
\multicolumn{3}{l}{$\Delta E^{(1)}$, $n_{\rm R}=2$}\\
\hline\hline
\multicolumn{2}{l|}{$n_{\rm T}$} & IV. & V. & VI. & VII. & Avg\\
\hline\hline
\multirow{4}{*}{$n_{\rm T}=3$}
& $\Delta E_{\rm eq}$ &10.55 &17.19 &22.16 &24.05 &18.49 \\
& $\Delta d_{\rm eq}$ &0.12 &0.14 &0.15 &0.15 &0.14 \\
&$\Delta\omega$ &-46.9 &-109.6 &-188.5 &-213.7 &-139.7 \\
&       IE         &13.01 &20.00 &27.43 &30.18 &22.65 \\
\hline
\multirow{4}{*}{$n_{\rm T}=4$}
& $\Delta E_{\rm eq}$ &10.76 &19.97 &23.77 &25.74 &20.06 \\
& $\Delta d_{\rm eq}$ &0.15 &0.20 &0.24 &0.21 &0.20 \\
&$\Delta\omega$ &-22.8 &-137.8 &-164.1 &-196.8 &-130.3 \\
&       IE         &13.19 &28.19 &36.84 &39.42 &29.41 \\
\hline\\
\multicolumn{3}{l}{$\Delta E^{(1)}$, $n_{\rm R}=3$}\\\hline
\multirow{4}{*}{$n_{\rm T}=2$}
& $\Delta E_{\rm eq}$ &32.90 &36.17 &41.62 &44.65 &38.84 \\
& $\Delta d_{\rm eq}$ &0.21 &0.23 &0.23 &0.22 &0.22 \\
&$\Delta\omega$ &-308.4 &-373.8 &-527.3 &-575.0 &-446.1 \\
&       IE         &47.26 &55.58 &65.66 &68.61 &59.28 \\
\hline
\multirow{4}{*}{$n_{\rm T}=4$}
& $\Delta E_{\rm eq}$ &1.52 &2.13 &2.93 &3.55 &2.53 \\
& $\Delta d_{\rm eq}$ &0.01 &0.02 &0.03 &0.03 &0.02 \\
&$\Delta\omega$ &-5.8 &-11.7 &-31.9 &-20.9 &-17.6 \\
&       IE         &1.45 &2.26 &2.83 &3.25 &2.45 \\
\hline\\
\multicolumn{3}{l}{$\Delta E^{(1)}$, $n_{\rm R}=4$}\\\hline
\multirow{4}{*}{$n_{\rm T}=2$}
& $\Delta E_{\rm eq}$ &38.08 &44.22 &54.03 &61.07 &49.35 \\
& $\Delta d_{\rm eq}$ &0.26 &0.31 &0.33 &0.33 &0.31 \\
&$\Delta\omega$ &-331.4 &-455.1 &-636.6 &-738.0 &-540.3 \\
&       IE         &60.39 &78.48 &106.16 &121.95 &91.75 \\
\hline
\multirow{4}{*}{$n_{\rm T}=3$}
& $\Delta E_{\rm eq}$ &1.26 &2.43 &3.81 &5.39 &3.22 \\
& $\Delta d_{\rm eq}$ &0.01 &0.02 &0.02 &0.04 &0.02 \\
&$\Delta\omega$ &-14.4 &-15.0 &-31.0 &-60.4 &-30.2 \\
&       IE         &1.16 &2.31 &3.56 &5.13 &3.04 \\
\hline\hline
\end{tabular}
\end{minipage}
\hspace{0.05\linewidth}
\begin{minipage}{0.45\textwidth}
\begin{tabular}{l>{\hspace{\thsp}}l|
>{\hspace{0.3pc}}r|
>{\hspace{\thsp}}r|
>{\hspace{\thsp}}r|
>{\hspace{0.3pc}}r||
>{\hspace{0.2pc}}r}
\multicolumn{3}{l}{$\Delta E^{(2)}$, $n_{\rm R}=2$}\\
\hline\hline
\multicolumn{2}{l|}{$n_{\rm T}$} & IV. & V. & VI. & VII. & Avg\\
\hline\hline
\multirow{4}{*}{$n_{\rm T}=3$}
& $\Delta E_{\rm eq}$ &18.92 &26.34 &31.39 &33.94 &27.65 \\
& $\Delta d_{\rm eq}$ &0.25 &0.27 &0.28 &0.26 &0.26 \\
&$\Delta\omega$ &-53.6 &-144.1 &-233.7 &-256.4 &-172.0 \\
&       IE         &32.34 &41.13 &52.19 &56.39 &45.51 \\
\hline
\multirow{4}{*}{$n_{\rm T}=4$}
& $\Delta E_{\rm eq}$ &18.14 &27.34 &30.60 &33.66 &27.43 \\
& $\Delta d_{\rm eq}$ &0.28 &0.34 &0.36 &0.34 &0.33 \\
&$\Delta\omega$ &-39.1 &-136.4 &-141.2 &-182.0 &-124.6 \\
&       IE         &32.30 &55.27 &67.82 &68.92 &56.08 \\
\hline\\
\multicolumn{3}{l}{$\Delta E^{(2)}$, $n_{\rm R}=3$}\\\hline
\multirow{4}{*}{$n_{\rm T}=2$}
& $\Delta E_{\rm eq}$ &12.33 &15.46 &19.08 &20.63 &16.88 \\
& $\Delta d_{\rm eq}$ &0.07 &0.08 &0.09 &0.08 &0.08 \\
&$\Delta\omega$ &-143.1 &-202.4 &-259.0 &-312.3 &-229.2 \\
&       IE         &12.95 &16.79 &21.26 &22.16 &18.29 \\
\hline
\multirow{4}{*}{$n_{\rm T}=4$}
& $\Delta E_{\rm eq}$ &4.66 &7.24 &9.33 &10.90 &8.03 \\
& $\Delta d_{\rm eq}$ &0.04 &0.08 &0.09 &0.09 &0.08 \\
&$\Delta\omega$ &-25.1 &-48.3 &-56.5 &-68.3 &-49.6 \\
&       IE         &4.83 &8.69 &10.69 &11.89 &9.03 \\
\hline\\
\multicolumn{3}{l}{$\Delta E^{(2)}$, $n_{\rm R}=4$}\\
\hline
\multirow{4}{*}{$n_{\rm T}=2$}
& $\Delta E_{\rm eq}$ &19.91 &22.19 &30.44 &33.75 &26.57 \\
& $\Delta d_{\rm eq}$ &0.12 &0.12 &0.16 &0.15 &0.14 \\
&$\Delta\omega$ &-211.2 &-276.1 &-418.1 &-487.4 &-348.2 \\
&       IE         &22.83 &25.10 &38.88 &42.45 &32.31 \\
\hline
\multirow{4}{*}{$n_{\rm T}=3$}
& $\Delta E_{\rm eq}$ &-1.30 &-3.64 &-5.32 &-6.49 &-4.19 \\
& $\Delta d_{\rm eq}$ &-0.01 &-0.03 &-0.03 &-0.04 &-0.03 \\
&$\Delta\omega$ &12.3 &12.4 &24.0 &51.1 &24.9 \\
&       IE         &1.43 &3.07 &4.39 &5.23 &3.53 \\
\hline\hline
\end{tabular}
\end{minipage}
\end{minipage}
\end{threeparttable}
\label{table:vertical_error}
\end{table}
\end{widetext}

\subsubsection{Alchemical predictions do not commute}
We note the asymmetry in the predictive power of first order based predictions
which is due to the lack of commutation: 
In general $\partial_\lambda E|_{\lambda = 0} \ne \partial_\lambda E|_{\lambda = 1}$,
except if reference and target Hamiltonian happen to differ only by translation, rotation, 
or parity (enantiomers, i.e. without accounting for parity violation). 
Within our restricted case of linear interpolations of iso-electronic systems, the
perturbing potential does differ only by sign. The integral over its product with the electron 
density, however, differs in general, 
i.e.~$\langle H_{\rm A} - H_{\rm B} \rangle_{\rm B} = \int d\fatr\; \rho_{\rm B} (v_{\rm A}-v_{\rm B})
\ne  \int d\fatr\; \rho_{\rm A} (v_{\rm A}-v_{\rm B}) = -\langle H_{\rm B} - H_{\rm A} \rangle_{\rm A}$.
As such, the error in estimating A based on B will not be the same as the error in estimating B based on A. 
Results in Table.~\ref{table:vertical_error} suggest that predictions downward the columns in the periodic table 
are more accurate than upward.
For example, predicting HBr using HF as a reference, a better
estimate is obtained (error = +25.7 kcal/mol) than for predicting HF 
using HBr as a reference (error = +61.1 kcal/mol).
Correspondingly, predicting HCl using HBr 
has an error = +5.4 kcal/mol, while the prediction of HBr using HCl has only an error of +3.6 kcal/mol.
Similar observations hold for bond lengths, and force constants. 
The asymmetry is also illustrated in Fig.~\ref{fig:p13d20}. 
$\Delta E(d, d_0)$ is not necessarily symmetric with respect to $\lambda=0.5$ for a given choice of $(d, d_0)$. 
Consequently, truncated Taylor series based predictions from either end will not be equally accurate. 

\subsubsection{Chemical accuracy}
\begin{figure}
\centering
\includegraphics[scale=0.4, angle=0, width=8.5cm]{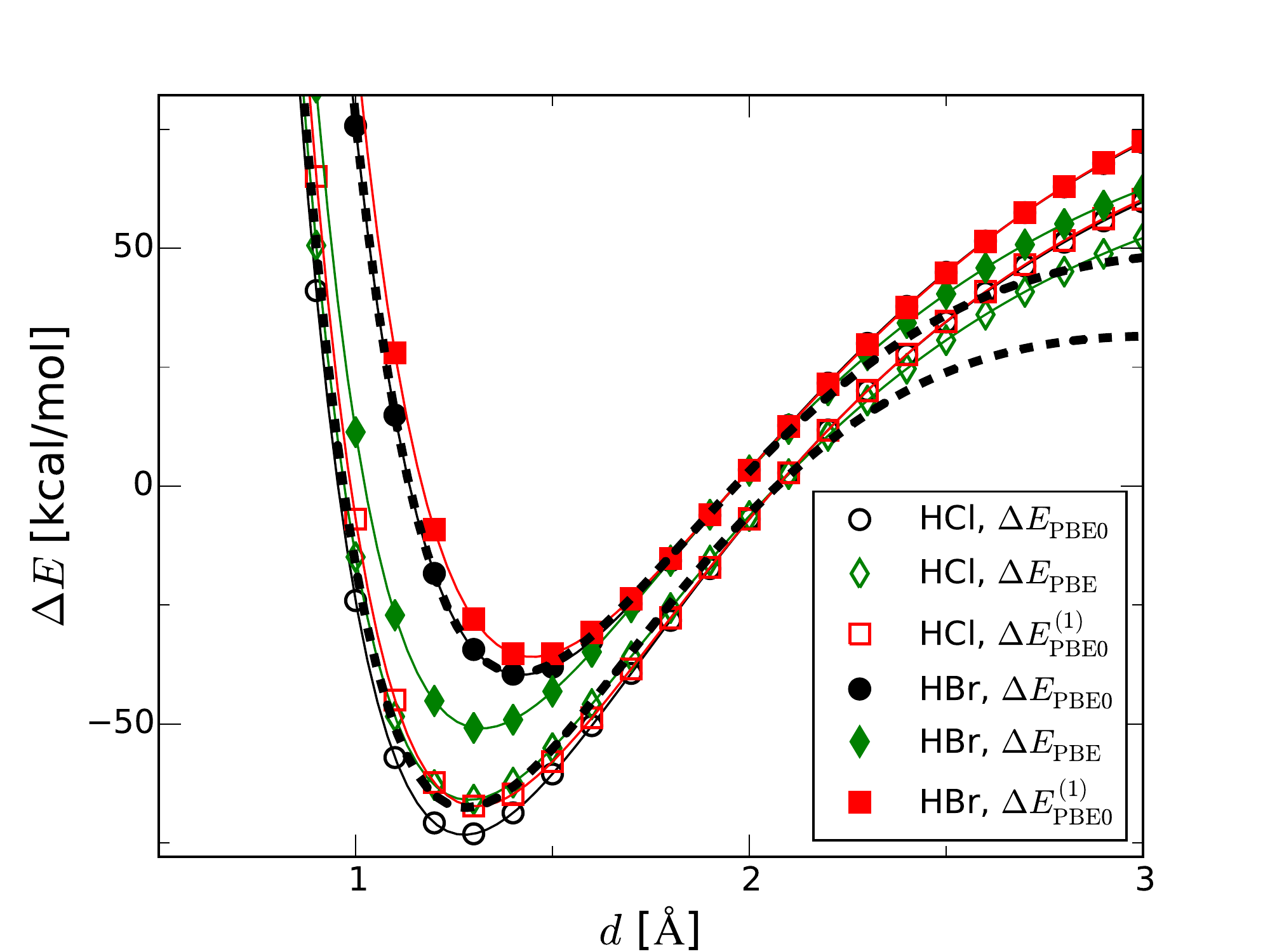}
\caption{
Alchemical predictions can be more accurate than approximated density functionals.  
Covalent binding potentials obtained from alchemical PBE0 estimate (red squares) 
and ordinary PBE (green diamonds) for HBr (full symbols) and HCl (empty symbols).
The alchemical estimate corresponds to Eq.~(\ref{eq:delta_E_explicit_1st}) using PBE0 density of HBr (HCl) in order to predict HCl (HBr).
For comparison, corresponding CCSD(T) results are shown as well (dashed).
}
\label{fig:pbe0vpbe}
\end{figure}

We have seen that very accurate, yet inexpensive, first order alchemical estimates can be made 
for vertical alchemical changes between third and fourth row elements
according to Eq.~(\ref{eq:d1E})---once the density is converged for a given reference molecule. 
Then, an interesting question is if the alchemical accuracy is on the same order of magnitude
as common approximations made when solving Schr\"odinger's equation.
We have investigated this point for alchemical coupling of HBr and HCl using hybrid 
and generalized gradient approximated DFT. 
When using PBE0\cite{PBE0} as the method for the reference compound, 
we find the first order based alchemical predictions according to Eq.~(\ref{eq:delta_E_explicit_1st}) 
to be in better agreement with the PBE0 results for the target compound 
than true generalized gradient based approximation PBE~\cite{BurkePerspective_2012}.
Fig.~\ref{fig:pbe0vpbe} illustrates this point for the covalent binding potentials of HCl and HBr calculated using 
PBE0, PBE0 based vertical first order alchemical predictions, and PBE. 
For all interatomic distances in the dissociative tail, 
the alchemical prediction (squares) is closer to PBE0 (circles) than PBE (diamonds). 
For the repulsive part of the potential, the alchemical prediction is substantially better
than PBE for HBr, and slightly worse than PBE for HCl.
For comparison we also included CCSD(T) results.  
These results amount to numerical evidence that 
the predictive power of vertical alchemical predictions can exceed the 
accuracy of common DFT approximations for third or forth row 
elements---if a sufficiently accurate electron density is provided for the reference compound.

\subsection{Vertical iso-valence-electronic changes involving single, double, and triple bonds}
\label{sec:heavy}

\subsubsection{Predicted potentials}
\begin{figure}[H]
\centering
\includegraphics[scale=0.4, angle=0, width=8.5cm]{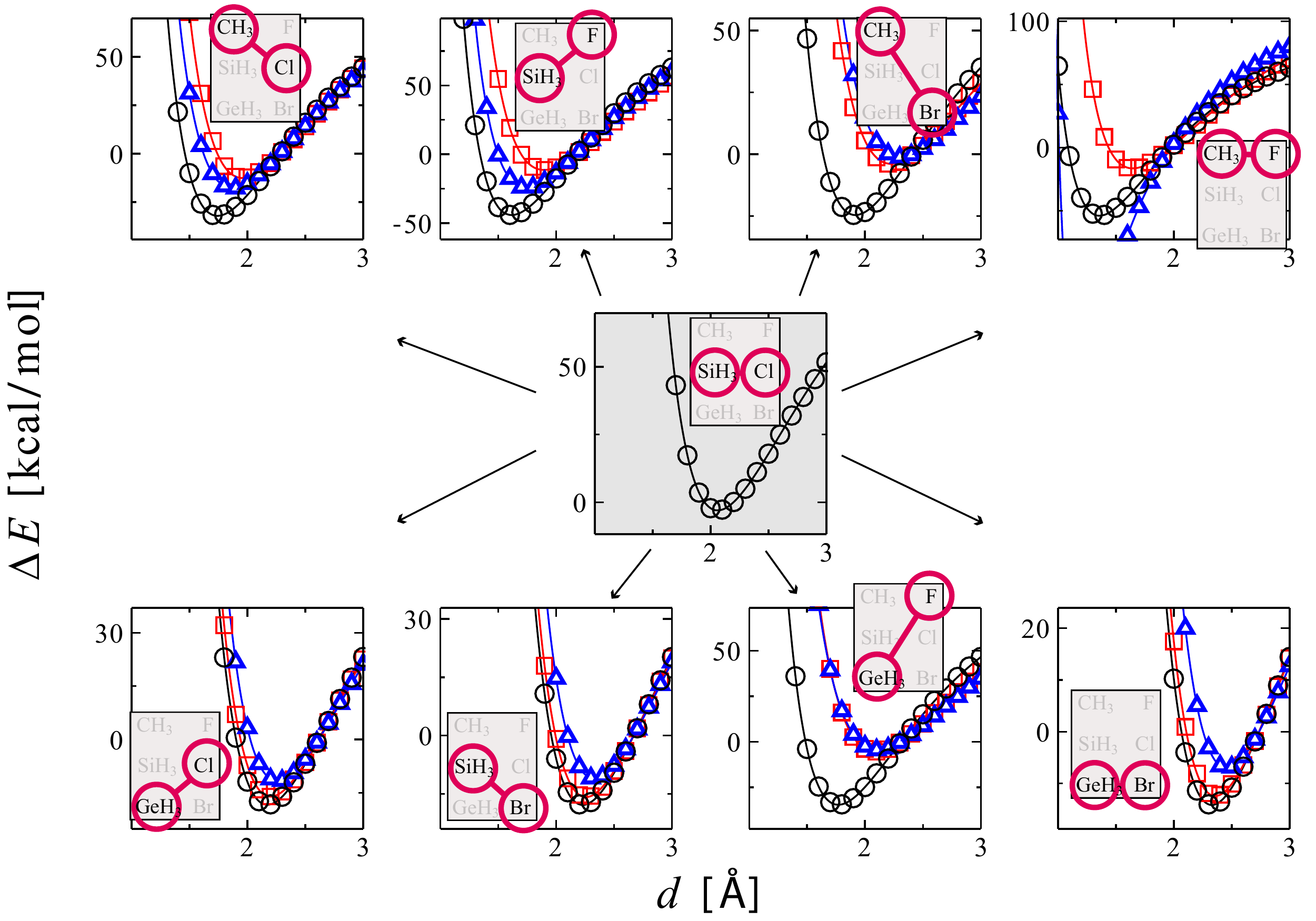}
\includegraphics[scale=0.4, angle=0, width=8.5cm]{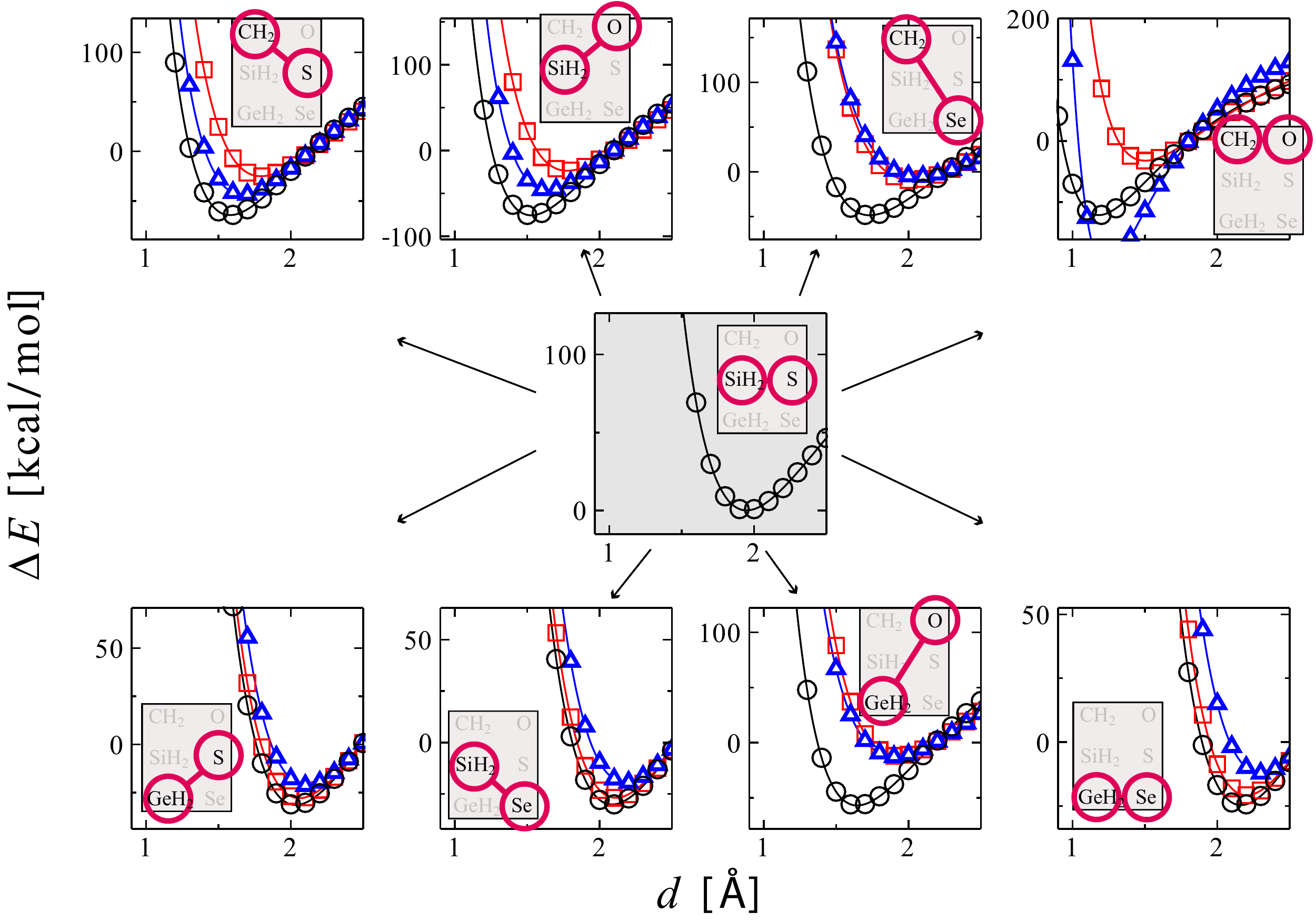}
\includegraphics[scale=0.4, angle=0, width=8.5cm]{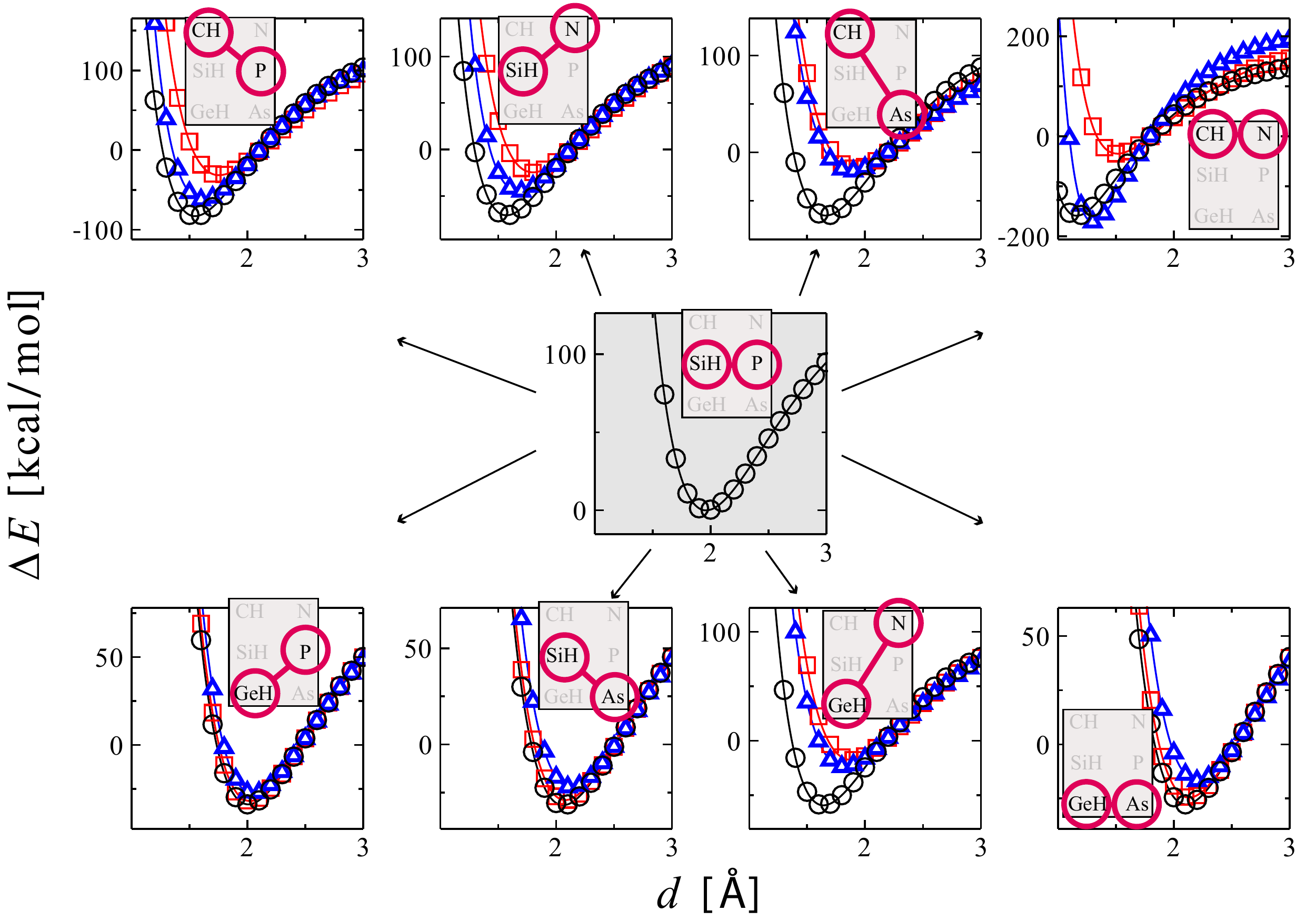}
\caption{
Alchemical predictions of single (top), double (middle), and triple (bottom) bond potentials.
Curves are shown for eight target systems (specified as insets), 
iso-electronic with reference molecule SiH$_3$Cl (upper panel), SiH$_2$S (middle panel), and HSiP (bottom panel)
True (black circles), first (red squares) and second (blue triangles) order vertical alchemical predictions 
of heavy atom bond dissociation curves.
}
\label{fig:heavy_fit}
\end{figure}

Having discussed covalent bonds involving hydrogen, 
we now turn to single (XH$_3${\tt -}Y), double (XH$_2${\tt =}Y), and triple (HX{\tt \#}Y) bonds among $p$-block elements.
Since third row elements can either be alchemically compressed to the corresponding second row ($n$ = 2) element in the same column,
or expanded to the fourth row ($n$ = 4) element, 
we chose third row ($n$ = 3) based reference systems for single, double, and triple bonds, 
namely SiH$_3$Cl, SiH$_2$S, and HSiP. 
The resulting eight alchemical paths are combinations of changing the Si atom (Si$\rightarrow$C, Si$\rightarrow$Ge)
or its binding partner (Cl$\rightarrow$F, Cl$\rightarrow$Br, S$\rightarrow$O, S$\rightarrow$Se, P$\rightarrow$N, P$\rightarrow$As).
In Figs.~\ref{fig:heavy_fit} first and second order alchemical predictions are shown for the bonding potential
using vertical transmutations from the three reference molecules.

More specifically, single bonds predictions have been investigated 
for making predictions using SiH$_3$Cl as a reference compound for 
the eight following molecules with 14 valence electrons:
CH$_3$F, CH$_3$Cl, CH$_3$Br ($n_{\rm X}=2$);
SiH$_3$F, SiH$_3$Br ($n_{\rm X}=3$);
and GeH$_3$F, GeH$_3$Cl, and GeH$_3$Br ($n_{\rm X}=4$).
For double bonds, we have considered predictions for the following eight unsaturated molecules 
12 valence electrons and using SiH$_2$S as a reference compound:
CH$_2$O, CH$_2$S, CH$_2$Se ($n_{\rm X}=2$);
SiH$_2$O, SiH$_2$Se ($n_{\rm X}=3$);
and GeH$_2$O, GeH$_2$S, and GeH$_2$Se ($n_{\rm X}=4$).
And finally for triple bonds, we have studied the following eight molecules with 10 valence electrons
and using HSiP as a reference compound:
HCN, HCP, HCAs ($n_{\rm X}=2$);
HSiN, HSiAs ($n_{\rm X}=3$);
and HGeN, HGeP, and HGeAs ($n_{\rm X}=4$).

Numerical results in Fig.~\ref{fig:heavy_fit} indicate qualitatively correct behavior for all predictions. 
Regarding quantitative performance, the accuracy of the alchemical prediction of $\Delta E(d, d_0)$
exhibits similar behavior as the one discussed above in the case of vertical changes in the hydrogen containing single bond: 
First order predictions (red) systematically achieve strong predictive power 
whenever the change involves the coupling of the third row element to a fourth row element. 
Corresponding second order predictions (blue) deteriorate the accuracy due to inflection points near $\lambda=0$. 
If the coupling involves one lighter element from the second row, the prediction is no longer quantitative. 
However, in these cases, second order predictions provide a slightly superior prediction.
If both atoms are simultaneously transmutated to lighter atoms from the second row, 
{\it e.g.} SiH$_3$Cl$\rightarrow$CH$_3$F, 
second order estimates over correct (change of sign) the first order prediction. 
In the case of one element transmutating upward the column, the other downward, 
the second order estimate is hardly distinguishable from the first order estimate.
We believe that the reason for this is that the coupling to the lighter element on the
one site in the molecule yields the concave behavior leading to an improvement in the prediction, 
while the coupling to the heavier element on the other site in the molecule yields the convex
behavior with the inflection point, leading to a deteriotation of the prediction. 
Effectively, these two effects cancel each other and result in the 
same predictive accuracy as the one obtained for the first order estimate.
This rationalization rests on the assumption that the discussion of Fig.~\ref{fig:p13d20} 
can be applied also to linear combination of effects at different transmutating sites.

\begin{figure}
\centering
\includegraphics[scale=0.4, angle=0, width=8.5cm]{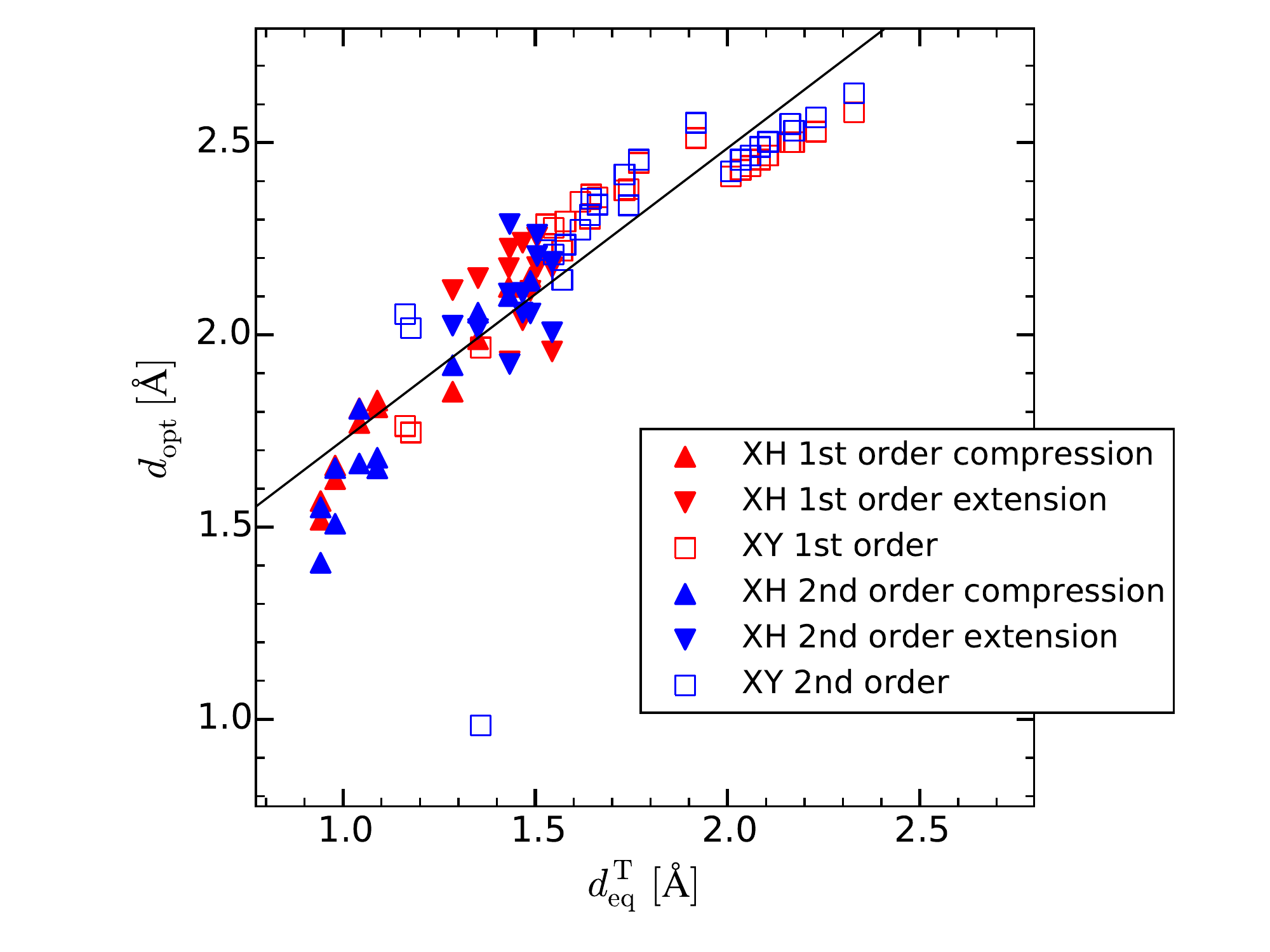}
\caption{Scatter plot of optimized reference bond length $d_{\rm opt}$ versus equilibrium bond length of target molecule, denoted by $d_{\rm eq}^{\rm T}$. Linear regression gives $d_{\rm opt} = 0.76\: d_{\rm eq} + 0.97$ \AA\ with MAE=0.11 \AA\ and RMSE=0.15 \AA.
First order (red empty squares) and second order (blue empty triangles) $d_{\rm opt}$
of covalent X-H bond stretching, as well as first order (red filled squares) 
and second order (blue filled triangles) $d_{\rm opt}$ of X{\tt -}Y, X{\tt =}Y, X{\tt \#}Y stretching are shown, 
where {\tt -}, {\tt =}, {\tt \#} stand for single, double, and triple bonds. Some of the alchemical paths are highlighted by black arrows.
All numbers are given in Tables I and II in the supplementary material.
}
\label{fig:d_scatter}
\end{figure}

\subsubsection{Integrated errors}
Above observations are consistent with the quantitative integrated prediction error measures (definitions in Sec.~\ref{sec:Error}) summarized in Table.~\ref{table:vertical_error_heavy}. 
All first order based predictions of target molecules implying a transmutation downward the periodic table
(columns 4/3, 3/4, 4/4) exhibit chemical accuracy with at most 1.83 kcal/mol deviation in minimal energy (GeH$_2$S), 
at most 0.04 {\AA} deviation in bond length (GeH$_2$Se),
at most -12.82 [cm$^{-1}$] deviation in wavenumber (SiH$_2$Se),
and at most 1.56 kcal/mol in integrated energy (GeHAs).
The best performance is achieved in the case of changing SiH$_3$Cl$\rightarrow$GeH$_3$Br 
with energy error $\Delta E =0.6$ kcal/mol and integrated $\mbox{IE}=0.9$ kcal/mol. 
Corresponding predictions of equilibrium distance deviates $0.03\:$\AA\ with vibration frequency deviate -1.1 cm$^{-1}$. 
First order predictions do not yield quantitative predictive power for changes involving lighter
elements (columns 2/3, 3/2, 2/4, 2/2, 4/2). 
The worst predictions are found for the simultaneous coupling to two lighter elements (colum 2/2) with
76.29 kcal/mol, 0.35 {\AA}, -568.76 cm$^{-1}$, and 41.45 kcal/mol deviation in minimum energy,
bond length, harmonic frequency, and integrated energy (HCN). 

While for all first order predictions all mutual deviations exhibit the same sign, 
second order corrections introduce the sign changes in minimum energy and bond length alluded to before, 
namely both third row elements couple to lighter elements from the second row (colum 2/2). 
Second order predictions for this column are even worse than the corresponding first order predictions.
Second order predictions only improve first order predictions in the case of 
columns 2/3 and 3/2, alas, not to a degree considered satisfying.

In summary, for changes 
corresponding to columns 4/3, 3/4, 4/4 first order 
based estimates yield chemical accuracy.
For changes corresponding to columns 2/2, 
first order based estimates are inaccurate but still better than second order estimates.
For changes corresponding to columns 2/4, and 4/2, 
first order based estimates are similar to second order estimates, yet both are inaccurate.
For changes corresponding to columns 2/3 and 3/2, 
second order based estimates are inaccurate but still better than first order estimates.

\subsection{Empirical $d_{\rm opt}$}
Above observations have been made for optimized $d_0$.
It should be noted that the choice of $d_0$ in Eq.~(\ref{eq:revE_lambda}) is crucial for 
linearizing the property of interest in alchemical coupling parameter $\lambda$, 
and hence essential for the performance of the perturbation based predictions.
We have found that the error minimizing $d_{\rm opt}$ has an approximately linear dependence
on the target molecule's equilibrium bond length $d_{\rm eq}$, 
no matter if the reference is the hydrogen containing single bond, or 
a single, double, or triple bond involving $p$-block elements from second, third, or fourth row.
Furthermore, the linear relationship is preserved, 
independent of the fact if predictions are made with first or second order estimates.
This relationship is shown in Fig.~\ref{fig:d_scatter}.
The parameters of a linear regression are specified as well. 
The outlier in Fig.~\ref{fig:d_scatter} at target $d_{\rm eq}\approx 1.35$\AA\ and
$d_{\rm opt}\approx 1.0$\AA\ is due to the second order prediction of 
SiH$_3$Cl$\rightarrow$CH$_3$F, i.e.~for the above discussed worst case scenario (column 2/2)
where a strong overcorrection has been found.

\begin{widetext}

\begin{table}
\centering 
\def\thsp{0.5cm}
\caption{Summary of error measure in Fig.~\ref{fig:heavy_fit} for first (upper table) and second order (lower table) based predictions. 
The reference molecules are in the left hand column as $H_{\rm R}=\{{\rm SiH}_3{\rm Cl},{\rm SiH}_2{\rm S},{\rm HSiP}\}$. 
The primary quantum number of the heavy atoms ${\rm X}=\{{\rm C}, {\rm Si}, {\rm Ge}\}$ and ${\rm Y}=\{{\rm N}, {\rm P}, {\rm As}, {\rm O}, {\rm S}, {\rm Se},{\rm F},{\rm Cl},{\rm Br}\}$ 
in target molecules XH$_3$-Y, XH$_2$=Y, and HX\#Y, respectively, 
are specified using the corresponding principal quantum numbers $n_{\rm X}$ and $n_{\rm Y}$ in each column. 
The right hand column ($n_{\rm X}$ = 4/$n_{\rm Y}$ = 4), for example, 
corresponds to predictions of molecules GeH$_3$Br, GeH$_2$Se, and HGeAs, respectively.
Error measures of each panel are collected in corresponding cell with unit [kcal/mol] 
for $\Delta E_{\rm eq}$ and IE, [\AA] for $\Delta d_{\rm eq}$, and $\Delta\omega$ is in [cm$^{-1}$].}
\begin{threeparttable}
\begin{tabular}{cl|
>{\hspace{\thsp}}r|
>{\hspace{\thsp}}r|
>{\hspace{\thsp}}r|
>{\hspace{0.1cm}}r|
>{\hspace{\thsp}}r|
>{\hspace{\thsp}}r|
>{\hspace{\thsp}}r|
>{\hspace{\thsp}}r
}
$\Delta E^{(1)}$\\
\hline\hline
\multicolumn{2}{l|}{$H_{\rm R}$} 
& 2/3 & 3/2 & 2/4 & 2/2 & 4/3 & 3/4 & 4/2 & 4/4 \\\hline
\multirow{4}{*}{\rotatebox[origin=c]{90}{SiH$_3$Cl}}
& $\Delta E_{\rm eq}$   &   9.04  &  18.25  &  10.37  &  20.37  &   0.95  &   0.80  &  15.83  &   0.60 \\
& $\Delta d_{\rm eq}$   &   0.22  &   0.27  &   0.31  &   0.29  &   0.03  &   0.03  &   0.33  &   0.03 \\
&$\Delta\omega$  & -91.83  &-169.02  &-106.27  &-237.85  & -12.05  &  -5.02  &-148.62  &  -1.09 \\
&     IE                &  10.14  &  16.82  &  12.98  &  14.85  &   1.18  &   0.98  &  16.74  &   0.88 \\
\hline
\multirow{4}{*}{\rotatebox[origin=c]{90}{SiH$_2$S}}
& $\Delta E_{\rm eq}$   &  21.69  &  30.78  &  22.87  &  52.34  &   1.83  &   1.28  &  26.77  &   1.42 \\
& $\Delta d_{\rm eq}$   &   0.21  &   0.25  &   0.29  &   0.32  &   0.02  &   0.03  &   0.28  &   0.04 \\
&$\Delta\omega$  &-191.24  &-241.71  &-201.69  &-436.06  &  -9.88  & -12.82  &-218.43  & -10.84 \\
&     IE                &  15.16  &  19.47  &  14.94  &  33.25  &   0.92  &   0.92  &  17.18  &   0.98 \\
\hline
\multirow{4}{*}{\rotatebox[origin=c]{90}{SiHP}}
& $\Delta E_{\rm eq}$   &  31.35  &  26.34  &  30.42  &  76.29  &   0.79  &   1.32  &  23.91  &   1.28 \\
& $\Delta d_{\rm eq}$   &   0.21  &   0.23  &   0.25  &   0.35  &   0.01  &   0.02  &   0.24  &   0.03 \\
&$\Delta\omega$  &-240.70  &-229.35  &-250.08  &-568.76  &  -1.77  &  -9.15  &-192.45  &  -2.34 \\
&     IE                &  22.28  &  19.42  &  23.73  &  41.45  &   0.93  &   1.22  &  18.35  &   1.56 \\
\hline\\
$\Delta E^{(2)}$\\
\hline
\multirow{4}{*}{\rotatebox[origin=c]{90}{SiH$_3$Cl}}
& $\Delta E_{\rm eq}$   &   7.79  &  11.84  &  13.29  &-103.39  &   3.00  &   2.54  &  17.60  &   2.68 \\
& $\Delta d_{\rm eq}$   &   0.13  &   0.13  &   0.39  &  -0.18  &   0.09  &   0.10  &   0.34  &   0.13 \\
&$\Delta\omega$  & -91.30  &-111.67  &-147.58  & 1912.55  & -22.96  & -18.27  &-176.49  & -13.20 \\
&     IE                &   6.48  &   8.47  &  16.19  &  32.26  &   3.36  &   3.61  &  17.63  &   3.83 \\
\hline
\multirow{4}{*}{\rotatebox[origin=c]{90}{SiH$_2$S}}
& $\Delta E_{\rm eq}$   &  12.46  &  17.88  &  28.09  &-101.05  &   5.18  &   4.59  &  29.36  &   5.36 \\
& $\Delta d_{\rm eq}$   &   0.09  &   0.11  &   0.33  &   0.06  &   0.08  &   0.09  &   0.26  &   0.13 \\
&$\Delta\omega$  &-113.32  &-128.03  &-258.05  & 1290.29  & -36.73  & -30.13  &-256.16  & -27.49 \\
&     IE                &   6.84  &   9.58  &  15.34  &  33.66  &   2.47  &   2.75  &  14.82  &   2.42 \\
\hline
\multirow{4}{*}{\rotatebox[origin=c]{90}{SiHP}}
& $\Delta E_{\rm eq}$   &  12.04  &  14.92  &  30.85  & -50.92  &   3.05  &   4.22  &  21.95  &   4.93 \\
& $\Delta d_{\rm eq}$   &   0.07  &   0.11  &   0.22  &   0.14  &   0.04  &   0.07  &   0.18  &   0.10 \\
&$\Delta\omega$  & -94.83  &-118.62  &-256.64  & 747.69  & -11.84  & -29.35  &-182.63  & -22.36 \\
&     IE                &   8.56  &   9.45  &  21.74  &  28.53  &   2.96  &   4.58  &  14.80  &   5.71 \\
\hline
\hline
\end{tabular}
\label{table:vertical_error_heavy}
\end{threeparttable}
\end{table}
\end{widetext}

\subsection{Non-vertical iso-electronic changes}
We are not aware of any mathematical limitation on how to construct alchemical coupling paths under isoelectronic condition. 
In addition to the investigation of predicted PES of iso-electronic compounds with the same geometry, as discussed in Sec.~\ref{sec:vertical} and Sec.~\ref{sec:heavy}, 
we have also investigated if one can use only {\em one} reference calculation in order to estimate 
the {\it entire} PES through ``non-vertical'' interpolations. 
In other words, we have also assessed the applicability of the Taylor expansion 
of Eq.~(\ref{eq:revE}) to non-vertical changes for
varying geometry and/or atom types and numbers between reference and target molecule. 

\subsubsection{Alchemical stretching of {\rm H$_{\it 2}^+$}}
\label{sec:h2p}
We now turn to the case of alchemical stretching of H$_2^+$ in order to understand the effect of 
varying geometry on alchemical predictions.
Since Hartree-Fock is numerically exact for one-electron systems, 
we have employed an atomic basis set in an "all-electron" (no PPs) 
calculation within the following alignment scheme: 
One proton is centered at $\fatR_1=(0,0,0)$, the other is aligned along the $+x$-axis. 
The reference system corresponds to H$_2^+$ at its equilibrium bond length. 
Stretching is accomplished not by pulling the atoms apart but rather 
by simultaneous annihilation and creation of nuclear charges at $\fatR_2^{\rm R}=(d_{\rm eq},0,0)$ 
and $\fatR_2^{\rm T}=(d,0,0)$, respectively. 
Once the SCF is done for $d_{\rm eq}$, the entire binding potential can be estimated up to $m = 4$ order, 
using Eq.~(\ref{eq:Model}), by scanning through various $d$, i.e.~setting $d_0=d_{\rm eq}$.

\begin{figure}
\centering
\includegraphics[scale=0.4, angle=0, width=8.5cm]{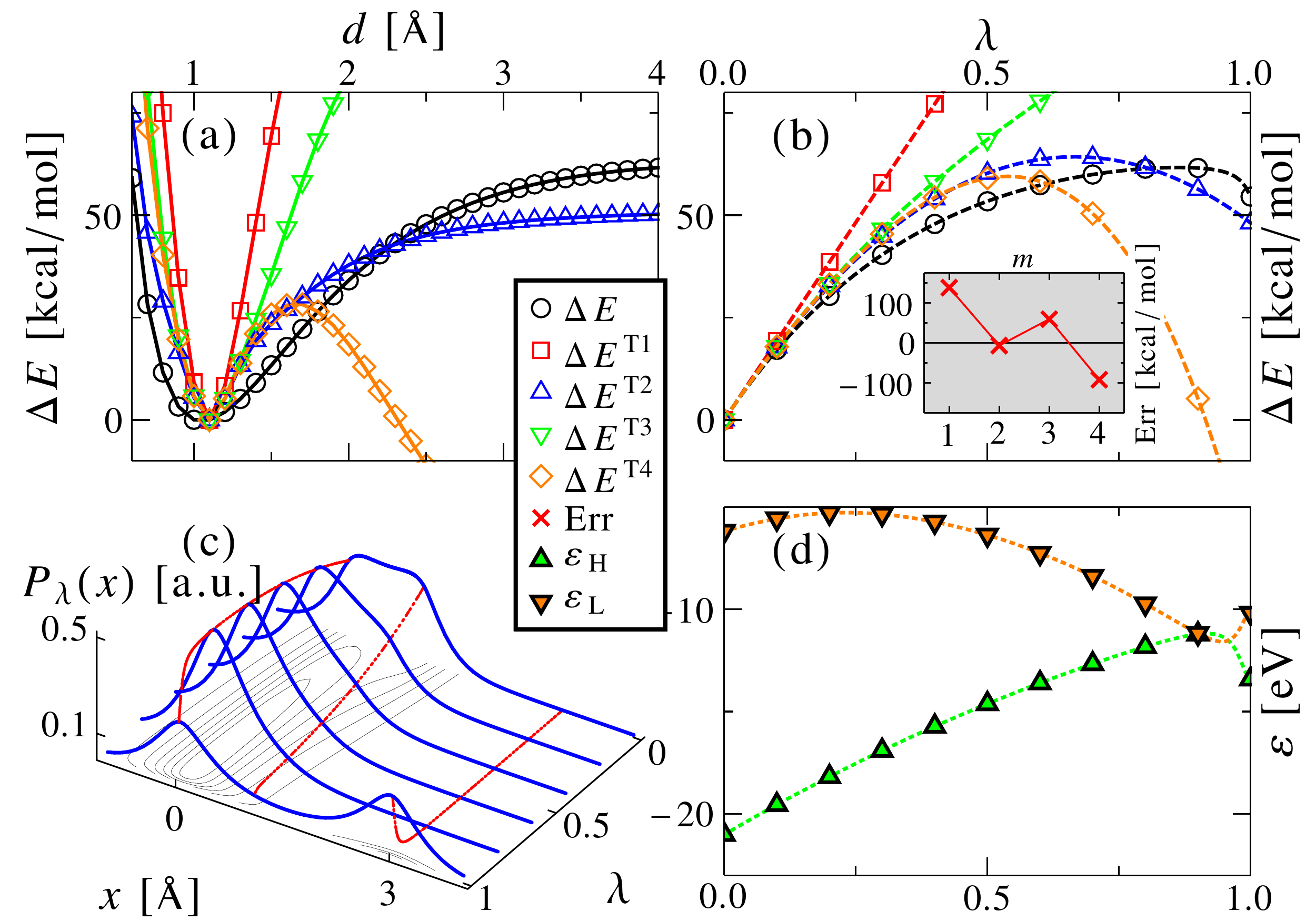}
\caption{
$m^{th}$ order truncated Taylor series of H$_2^+$ are denoted by $\Delta E^{(m)}$ in (a) as a function of $d$ at $\lambda=1$ and in (b) as a function of $\lambda$ at $d=3\:$\AA. Inset shows the error of $\Delta E^{(m)}$ at $\lambda=1$ (c) Integrated density $P_\lambda(x) = \int dydz\:\rho_\lambda(\fatr)$ where $\fatr=(x,y,z)$, is presented as a function of both $\lambda$ and $x$ at $d = 3$\:\AA\, where the integrated density values at nuclei locations highlighted by red lines at $x=0\:$\AA, $x=1.1\:$\AA, and $x=3\:$\AA, while contour lines are draw at the bottom. (d) HOMO/LUMO levels, denoted by $\varepsilon_{\rm H}$ and $\varepsilon_{\rm L}$ respectively, are plotted as a function of $\lambda$ at $d=3\:$\AA.
}
\label{fig:h2p}
\end{figure}

Results are shown in Fig.~\ref{fig:h2p}(a). 
Due to the variational principle for linearly coupled alchemical Hamiltonians,\cite{Anatole_ijqc_2013} 
$\Delta E^{(1)} > \Delta E$ for all interatomic distances. 
Inclusion of second order term improves upon the first order prediction, yielding a reasonable binding potential. 
However, when including third and fourth order the performance deteriorates again with oscillating behaviour 
for varying order (Fig.~\ref{fig:h2p}(a) and inset of (b)), as $\Delta E^{(3)}$ overshoot and $\Delta E^{(4)}$ over corrects. 
Overall $\Delta E^{(2)}$ gives the best prediction. 

To explain the oscillating behaviour in Taylor expansion order, 
we investigate in more detail how the system responds to alchemical perturbation. 
When $\lambda$ increases gradually from 0 to 1, the nuclear charge decreases from 1 to 0 at $\fatR_2^{\rm R}$, 
while increasing from 0 to 1 at $\fatR_2^{\rm T}$. 
Using the alchemical derivatives at $\lambda = 0$, truncated Taylor series based estimates are plotted 
along the true energy 
in Fig.~\ref{fig:h2p}(b) as a function of $\lambda$ at $d=3\:$\AA.
$\Delta E^{(1)}$, $\Delta E^{(2)}$, $\Delta E^{(3)}$, and $\Delta E^{(4)}$ are linear, quadratic, third order, and fourth order polynomials, respectively.
Clearly, the truncated Taylor series will fail to converge to $\Delta E$ at $\lambda=1$ due to a 
sharp change of $\Delta E$ at $\lambda\approx 0.9$. 
This implies a strong nonlinear electronic response occurring late in the alchemical coupling regime, 
resulting in the oscillating behaviour of the predicted PES in Figs.~\ref{fig:h2p}(a) and (b). 
Note that while the sign of error alternates, the magnitude of error also increases as one increases the order. 
Similar behaviour can be observed for other values of $d$.

The energy gain starting at $\lambda\approx 0.9$ is due to a rapid rearrangement of electron density for $\lambda > 0.9$. 
This is illustrated in Fig.~\ref{fig:h2p}(c) where the integrated electron density $P_\lambda(x)$ 
is plotted as a function of both $\lambda$ and $x$ at $d=3\:$\AA. 
Cohen and Mori-S{\'a}nchez already pointed out for H$_2^+$ the dramatic changes in electronic structure 
for infinitesimally small changes in nuclear charges at infinite distance.\cite{AJCohen_JCP_2014} 
One would expect this effect to intensify as more basis functions are taken into account. 
This behaviour can be seen in Fig.~\ref{fig:h2p}(c). 
The locations of the proton at origin $\fatR_1$, 
as well as the location of the annihilated proton at $\fatR_2^{\rm R}$, and created proton at $\fatR_2^{\rm T}$, are indicated by red lines. 

Further analysis shows that for $\lambda>0.5$, both ground and first excited state orbitals are localized: 
The electronic ground state is localized at $\fatR_1$ while the first excited state is localized at $\fatR_2^{\rm T}$. 
At $\lambda\approx 0.9$, the two eigenvalues become degenerate, 
resulting in a rapid change of the ground state density in order to meet the non polar symmetry requirement of H$_2^+$, 
by taking a linear combination of both ground and first excited state. 
Note that there is no orbital node at midpoint, indicating a true ground state 
for a dissociated H$_2^+$ molecule. 
The degeneracy occurs for the system with fractional nuclear charges at $\lambda\approx 0.9$. 
The dramatic change in density stabilizes the system in $\lambda$, giving rise to the sharp decrease in energy in Fig.~\ref{fig:h2p}(b), as $\lambda$ increases from 0.8 to 1.
Perturbation theory for degenerate cases might be necessary to properly account for this case.
The degeneracy of the ground state and first excited state is shown for the eigenvalue crossing 
in Fig.~\ref{fig:h2p}(d): 
The eigenvalue of the (highest) occupied molecular orbital (HOMO) and 
lowest unoccupied molecular orbital (LUMO) are plotted as a function of $\lambda$. 
The degeneracy breaks when ground state and first excited state switch order, which results in a delocalized ground state. 
By contrast, note that the eigenvalues will not cross each other if the stretching is carried out by moving $\fatR_2^{\rm R}$ in real space. 

Crossing of eigenvalue surfaces limits the radius of convergence of alchemical Taylor expansion series within electronic ground-state theories. 
As a result, the Taylor expansion for this system is not convergent at $\lambda = 1$, 
similar to well known cases in M{\o}ller-Ploesset theory.\cite{MP2_criticalPoint, MP2_diverge, MP2_crossingIssues} 
For asymmetric alchemical interpolations, as exemplified for the following examples in this study, 
as well as in previous studies,\cite{Anatole_prl_2005, Anatole_jcp_2006, Anatole_jcp_2009, CHIMIA_2014} the energy is typically smooth in all $\lambda$ values, and derivative based expansions are expected to converge. 


\subsubsection{Non-vertical iso-electronic changes in ten electron systems}
\label{sec:non-vertical}
In the final section of this paper, we consider alchemical non-vertical changes of molecules with ten electrons.
More specifically, we present numerical results of non-vertical iso-electronic changes 
involving bond stretching in second row systems \{CH$_4$, NH$_3$, H$_2$O, HF\}, using all electron DFT. 
The H$_2^+$ example has indicated that non-vertical changes can profit from second order estimates. 
Since exact analytical expressions are not available for systems with so many electrons, and
since no PPs are involved, we have relied on approximative second order expressions IPA and CP,
rather than on finite difference expressions (see Methods section above).

\begin{figure}
\centering
\includegraphics[scale=0.4, angle=0, width=8.5cm]{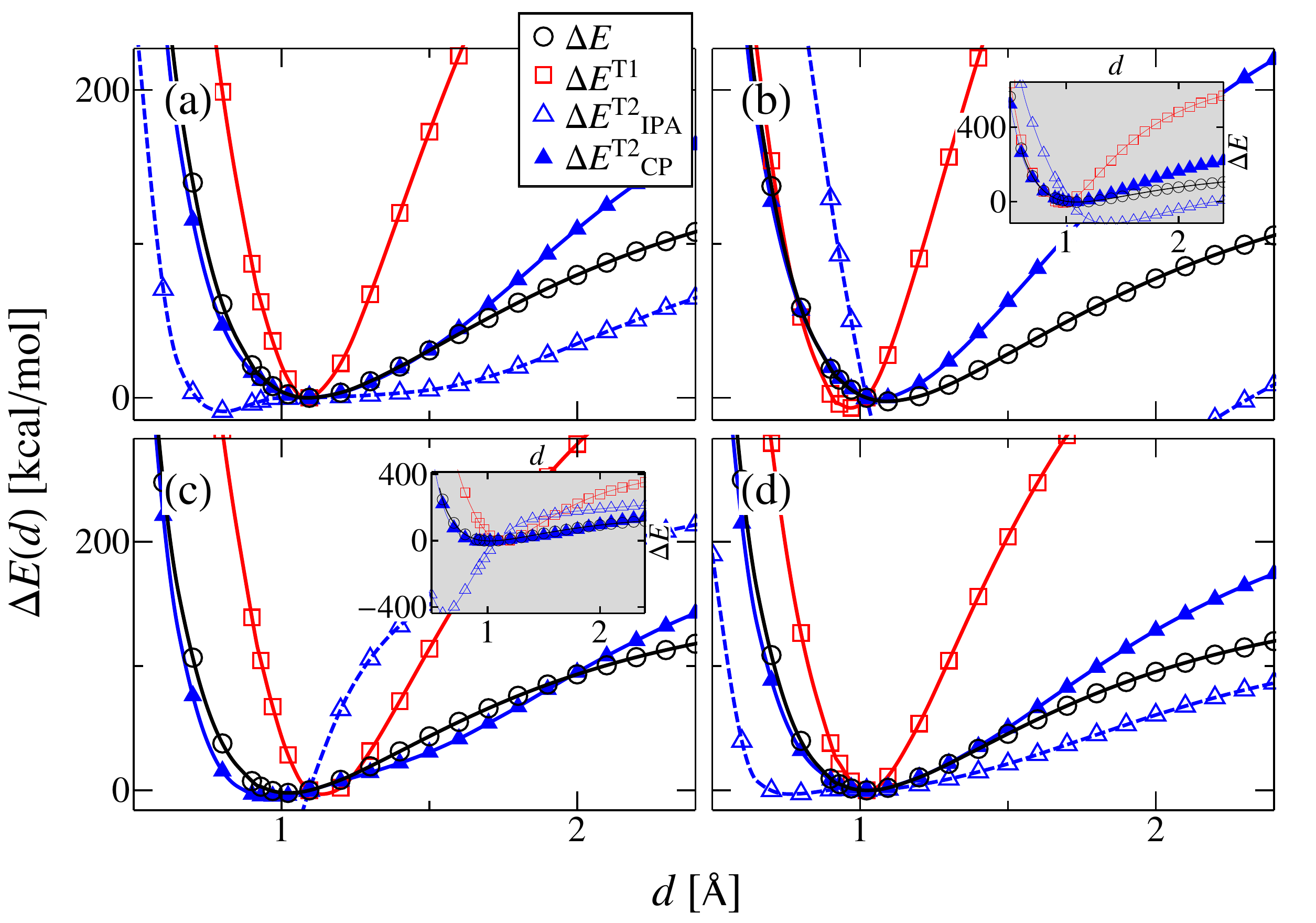}
\caption{Energy difference $\Delta E$, first order truncated Taylor series $\Delta E^{(1)}$, second order truncated Taylor series calculated by coupled perturbed $\Delta E^{(2)}_{\rm CP}$, and second order truncated Taylor series calculated by independent particle approximation $\Delta E^{(2)}_{\rm IPA}$ are plotted as black circles, red squares, blue open triangles, and blue filled triangles respectively. Coupling Hamiltonians are arranged as follow: (a) CH$_4\rightarrow$\:CH$_4$, (b) NH$_3\rightarrow$\:CH$_4$, (c) CH$_4\rightarrow$\:NH$_3$, and (d) NH$_3\rightarrow$\:NH$_3$. Insets of (b) and (c) show the zoom-out energy scale for overall landscape.}
\label{fig:ch4Stretch-2x2}
\end{figure}

\begin{table}
\centering 
\def\thsp{0.8pc}
\caption{Prediction errors using first and second order based alchemical estimates for non-vertical changes in ten electron systems.
Deviation of predicted energy minimum from actual $\Delta E_{\rm eq}$ [kcal/mol], corresponding bond length deviation $\Delta d_{\rm eq}$ [\AA], and vibration frequency $\Delta\omega{\rm eq}$. Numerical results from first order $\Delta E^{(1)}$, second order with independent particle approximation $\Delta E^{(2)}_{\rm IPA}$, and second order with coupled perturbed $\Delta E^{(2)}_{\rm CP}$ truncated Taylor series are presented. Average is calculated for each row in Avg column.}
\begin{threeparttable}
\begin{tabular}{ll|
>{\hspace{\thsp}}r|
>{\hspace{\thsp}}r|
>{\hspace{0.4pc}}r|
>{\hspace{0.3pc}}r|
|>{\hspace{0.4pc}}r
}
\multicolumn{2}{l}{$\Delta E^{(1)}$:}\\\hline\hline
\multicolumn{2}{l|}{$H_{\rm R}$}
& CH$_4$ & NH$_3$ & H$_2$O & HF & Avg\\\hline\hline
\multirow{3}{*}{CH$_4$}
&$\Delta E_{\rm eq}$& -0.22 & -2.30 & -9.63 & -21.99 & -8.54 \\
&$\Delta d_{\rm eq}$& -0.01 & 0.12 & 0.22 & 0.30 & 0.15 \\
&$\Delta\omega$& 6061. & 3545. & 2641. & 1874. & 3530. \\
\hline
\multirow{3}{*}{NH$_3$}
&$\Delta E_{\rm eq}$& -4.33 & -0.04 & -4.34 & -14.01 & -5.68 \\
&$\Delta d_{\rm eq}$& -0.13 & 0.002 & 0.10 & 0.19 & 0.04 \\
&$\Delta\omega$& 4218. & 4268. & 4622. & 2449. & 3889. \\
\hline
\multirow{3}{*}{H$_2$O}
&$\Delta E_{\rm eq}$& -32.78 & -5.97 & -0.0000 & -4.88 & -10.91 \\
&$\Delta d_{\rm eq}$& -0.27 & -0.11 & 0.0005 & 0.09 & -0.07 \\
&$\Delta\omega$& 4360. & 3933. & 3886. & 2568. & 3687. \\
\hline
\multirow{3}{*}{HF}
&$\Delta E_{\rm eq}$& -120.8 & -38.63 & -7.95 & 0.0003 & -41.85 \\
&$\Delta d_{\rm eq}$& -0.44 & -0.25 & -0.10 & -0.0008 & -0.20 \\
&$\Delta\omega$& 5772. & 4047. & 3982. & 3560. & 4340. \\
\hline
\\\multicolumn{2}{l}{$\Delta E^{(2)}_{\rm IPA}$:}\\
\hline
\multirow{3}{*}{CH$_4$}
&$\Delta E_{\rm eq}$& -8.99 & -437.5 & -935.2 & -1356. & -684.4 \\
&$\Delta d_{\rm eq}$& -0.31 & -0.41 & -0.40 & -0.39 & -0.38 \\
&$\Delta \omega$ & 2207. & 9625. & 12610. & 14600. & 9761. \\
\hline
\multirow{3}{*}{NH$_3$}
&$\Delta E_{\rm eq}$& -113.4 & -4.10 & -673.2 & -1394. & -546.1 \\
&$\Delta d_{\rm eq}$& 0.29 & -0.27 & -0.45 & -0.43 & -0.21 \\
&$\Delta \omega$ & 470.9 & 1471. & 11200. & 12660. & 6452. \\
\hline
\multirow{3}{*}{H$_2$O}
&$\Delta E_{\rm eq}$& -198.7 & -90.18 & -0.007 & -809.4 & -274.6 \\
&$\Delta d_{\rm eq}$& 0.16 & 0.21 & 0.007 & -0.43 & -0.01 \\
&$\Delta \omega$ & 2949. & 1124. & -1484. & 9157. & 2937. \\
\hline
\multirow{3}{*}{HF}
&$\Delta E_{\rm eq}$& -212.6 & -140.4 & -64.95 & 0.002 & -104.5 \\
&$\Delta d_{\rm eq}$& 0.05 & 0.11 & 0.15 & -0.001 & 0.08 \\
&$\Delta\omega$& 6035. & 4110. & 1975. & -1177. & 2736. \\
\hline
\\\multicolumn{2}{l}{$\Delta E^{(2)}_{\rm CP}$:}\\
\hline
\multirow{3}{*}{CH$_4$}
&$\Delta E_{\rm eq}$& -0.39 & -2.74 & -19.70 & -41.12 & -15.99 \\
&$\Delta d_{\rm eq}$& -0.02 & -0.07 & -0.10 & -0.10 & -0.07 \\
&$\Delta\omega$& 1663. & -495.7 & 2116. & 3584. & 1717. \\
\hline
\multirow{3}{*}{NH$_3$}
&$\Delta E_{\rm eq}$& 0.90 & -0.04 & -1.84 & -15.17 & -4.04 \\
&$\Delta d_{\rm eq}$& -0.04 & 0.007 & -0.06 & -0.10 & -0.05 \\
&$\Delta\omega$& 2133. & 140.4 & -1480. & 1690. & 620.9 \\
\hline
\multirow{3}{*}{H$_2$O}
&$\Delta E_{\rm eq}$& 5.78 & 0.77 & 0.0000 & -1.51 & 1.26 \\
&$\Delta d_{\rm eq}$& -0.08 & -0.02 & 0.002 & -0.04 & -0.03 \\
&$\Delta\omega$& 2207. & -75.3 & -194.2 & -345.3 & 398.0 \\
\hline
\multirow{3}{*}{HF}
&$\Delta E_{\rm eq}$& 13.12 & 4.25 & 0.65 & 0.001 & 4.50 \\
&$\Delta d_{\rm eq}$& -0.14 & -0.07 & -0.02 & -0.0006 & -0.06 \\
&$\Delta\omega$& 3501. & 2272. & 1074. & -247.5 & 1650. \\
\hline\hline
\end{tabular}
\label{table:ch4-hf_EdOPT} 
\end{threeparttable}
\end{table}

\subsubsection{Predicted potentials}

Fig.~\ref{fig:ch4Stretch-2x2} illustrates the prediction of R-H covalent bond potentials for CH$_4$ and NH$_3$, 
predicted from alchemical derivatives using the electronic structure obtained by a single SCF. 
As a reference system we used once the relaxed CH$_4$ system \big(panels (a), (c)\big),
and once the relaxed NH$_3$ \big(panels (b), (d)\big) geometry. 
For the chemical composition of $H_{\rm R}$ being the same as $H_{\rm T}$ and only the bond
being stretched (Fig.~\ref{fig:ch4Stretch-2x2}(a), (d)), 
the first order estimate constitutes an upper bound, {\it i.e.} it always overshoots 
due to the concave behaviour of $\Delta E$ as a function of $\lambda$, also on display in Fig.~\ref{fig:h2p}(b).
When also changing the chemical compositions from CH$_4$ and NH$_3$ or vice versa, 
the first order estimate does not even capture the changes in equilibrium bond length 
(Fig.~\ref{fig:ch4Stretch-2x2}(b) and (c)). 

$\Delta E_{\rm IPA}^{(2)}$ yields a saddle point in Fig.~\ref{fig:ch4Stretch-2x2}(a) and (d), 
instead of a minimum at optimized geometry. 
When the chemical compositions of $H_{\rm R}$ and $H_{\rm T}$ are different, 
$\Delta E_{\rm IPA}^{(2)}$ results in in dramatic errors (worse than first order erstimates), 
as shown in the energy zoom out in the insets of Fig.~\ref{fig:ch4Stretch-2x2}(b) and (c). 
The poor predictivey power of IPA has also recently been pointed out by Pulay and co-workers~\cite{Pulay2016}.
By contrast, $\Delta E_{\rm CP}^{(2)}$ yields a very reasonable binding potential, 
albeit still far from being chemically accurate. 
The superior performance of $\Delta E_{\rm CP}^{(2)}$, with respect to $\Delta E_{\rm IPA}^{(2)}$, indicates that the contributions of Coulomb and xc energy due to density response are crucial. 
In other words, matrix elements $\mathbf{J}_{ia,jb}$ and $\mathbf{X}_{ia,jb}$ 
in Eq.~(\ref{eq:dn_cpks_matrixElements}) should not be neglected for non-vertical alchemical perturbations. 

\begin{figure}
\centering
\includegraphics[scale=0.4, angle=0, width=8.5cm]{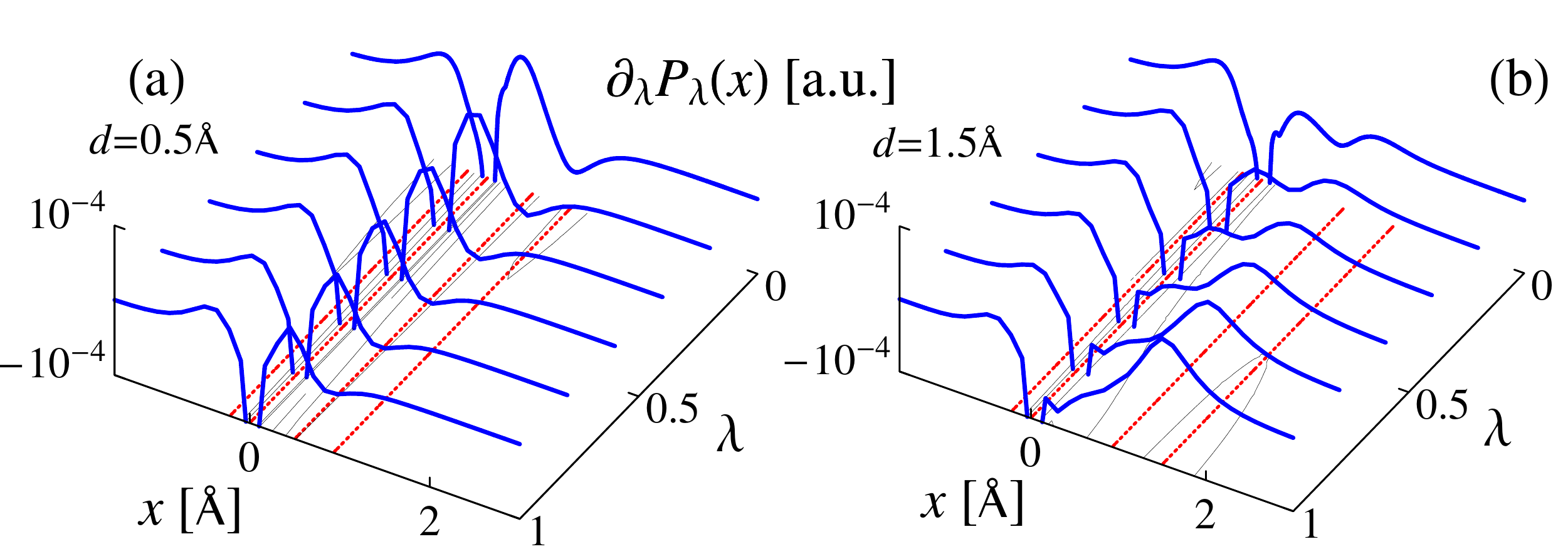}
\caption{
$\partial_\lambda P_\lambda(x)$ is calculated by finite difference $\partial_\lambda P_\lambda(x)\approx\frac{P_{\Delta\lambda}(x)-P_{\lambda=0}(x)}{\Delta\lambda}$.
$\partial_\lambda P_\lambda(x)$, of HF$\rightarrow$H$_2$O at (a) $d=0.5\:$\AA\ for compression and at (b) $d=1.5$\:\AA\ for extension are plotted as a function of $x$ and $\lambda$. Nuclear positions are highlighted by red dashed lines, with F$\rightarrow$O at $x=0\:$\AA, H$\rightarrow$void at $x=0.93\:$\AA, two void$\rightarrow$H at $x=-0.22\:$\AA\ and $x=d$, where void denotes the nuclei with zero charge.}
\label{fig:hf-h2o_density}
\end{figure}

Different predictive accuracy is found for compressing bonds $d<d_{\rm eq}$ versus stretching bonds $d>d_{\rm eq}$.
$\Delta E^{(2)}_{\rm CP}$ performs better in the region $0.5\:{\rm \AA}\leq d\leq 1.5\:{\rm \AA}$. 
Similar behaviour is also observed for other alchemical paths of compressing vs stretching bond.
Also in this case, the aforementioned non-commutative asymmetric behavior of the predictions is observed. 
Namely, the $\Delta E^{(2)}_{\rm CP}$ based prediction for ${\rm CH_4}\rightarrow {\rm NH_3}$ in Fig.~\ref{fig:ch4Stretch-2x2}(c) is more accurate than for ${\rm NH_3}\rightarrow {\rm CH_4}$ in Fig.~\ref{fig:ch4Stretch-2x2}(b).
Note that abrupt changes in electronic structure, as observed for H$_2^+$ in Sec.~\ref{sec:h2p}, 
are not present when coupling these systems.\cite{WYang_prl_2008, WYang_cr_2012}
Since the accuracy of the second order estimate is determined by how linearly the electron density 
rearranges as a function of $\lambda$, one expects a near-constant $\partial_\lambda\rho$ 
for negligible higher order contributions. 
This is confirmed through inspection of the integrated density response of the alchemical path 
HF$\rightarrow$H$_2$O in Fig.~\ref{fig:hf-h2o_density}.
$\partial_\lambda P_\lambda(x)$ varies less when $\lambda$ changes from zero to one for $d=0.5\:$\AA\ Fig.~\ref{fig:hf-h2o_density}(a), when compared with $d=1.5\:$\AA\ Fig.~\ref{fig:hf-h2o_density}(b). 
A near constant $\partial_\lambda P_\lambda(x)$ at $d=0.5\:$\AA\ results in improved predictive accuracy.

\subsubsection{Integrated errors}

Table.~\ref{table:ch4-hf_EdOPT} summarizes the results for all 4$\times$4 combinations of $H_{\rm R}\rightarrow H_{\rm T}$, 
where mutual predictions of covalent bond potentials in $H_{\rm T}$: \{CH$_4$, NH$_3$, H$_2$O, HF\} 
are obtained based on only the single point wavefunctions obtained for the relaxed geometry 
of $H_{\rm R}$: \{CH$_4$, NH$_3$, H$_2$O, HF\}, respectively. 
This coupling matrix in chemical space (Table.~\ref{table:ch4-hf_EdOPT}) is not symmetric (due to the non-commutative 
properties discussed above). 
Off-diagonal elements correspond to coupling paths involving changes in chemical composition {\em and} geometry. 
Diagonal elements correspond to coupling paths that involve only changes in geometry, i.e.~for the same stoichiometry. 
Note that all error measures have been obtained via cubic spline fits. 
Therefore, also the predicted $\Delta E$ and the location of the energy minimum
can be slightly non-zero even for the diagonal elements. 
These values should be considered noise: 
For the diagonal elements only the harmonic frequencies are meaningful.

Table.~\ref{table:ch4-hf_EdOPT} confirms the trends observed above for first and second order. 
In general, best predictive power is found when the chemical composition of 
$H_{\rm R}$ is the same as $H_{\rm T}$ (diagonal elements). 
When the chemical composition of $H_{\rm T}$ differs from $H_{\rm R}$ the predictive accuracy deteriorates.
This is not surprising and due to the perturbing Coulomb potential being placed on the heavy atom 
in order to mutate it, e.g.~from carbon to fluorine. 
Because of the strong accumulation of electron density (cusps) at the heavy atom's site (6 to 7 electrons for
carbon to fluorine, respectively), this perturbation is quite severe. 
In the case of the diagonal element, by contrast, only the hydrogen atom is being annihilated and created, 
implying that the perturbing potential acts on the hydrogen atom's electronic density which is
built up by only 1 electron. 
This implies a less severe perturbation, and therefore worse predictive power can be
expected for off-diagonal elements.

The crucial importance of Coulomb and xc energy contribution to density response for second order alchemical perturbation 
is also confirmed for the other cases in Table.~\ref{table:ch4-hf_EdOPT}. 
These results clearly underscore the observation that IPA is a (very) poor approximation when it comes to estimate alchemical changes, yielding even worse predictions than the first order estimates.
Interestingly enough, the first order estimate is even competitive in comparison to the second order CP predictions. 
For example, using CH$_4$ as a reference compound the first 
order prediction deviates on average by -8.54 kcal/mol in the energy, 
while $\Delta E^{(2)}_{\rm CP}$ deviates -15.99\:kcal/mol. 
However, as the reference compound moves to the right hand side of the periodic table, 
the second order CP based estimate becomes more accurate than the first order based estimate.

An additional aspect can be confirmed from inspection of Table.~\ref{table:ch4-hf_EdOPT}: 
The larger the perturbing potential, the worse the predictive accuracy of derivative based estimate.
More specifically, the larger the integrated norm of the difference 
between reference and target potential in the electronic Hamiltonian, the worse the predictive power. 
For example, using CH$_4$ as a reference compound, the prediction will be increasingly
worse in the order of the respective predictions for NH$_3$, H$_2$O, and HF.
Conversely, using HF as a reference compound, the prediction will be increasingly
worse in the order of the respective predictions for 
H$_2$O, NH$_3$, and CH$_4$.
Note that this is true for all first as well as second order estimates. 

\section{Conclusions}
\label{sec:conclusions}
The performance of truncated Taylor series for predicting alchemical vertical changes
in covalent bonding has been investigated in iso-electronic 
chemical spaces spanned by the external potentials of small molecules. 
For vertical linear transmutations (same geometry, same number of atoms, differing nuclear charges) 
our results suggest that chemical accuracy is possible
when interpolating molecules containing $p$-block atoms 
from the third and fourth row 
using first order (Hellmann-Feynman theory) based predictions. 
Since first order estimates are analytical, 
this finding implies that one can scan potential energy surfaces 
of very many molecules with unprecedented accuracy and speed 
as long as their stoichiometries are restricted to 
third and fourth row main group chemistries.
First order based predictions of chemistries involving
second row elements are only correct to a degree considered qualitative.

Overall, we have found second order estimates to not provide sufficient
improvement with respect to first order predictions (often even worse results) 
to warrant the investment in the additional overhead incurred.
First order estimates are more accurate not because higher order terms are negligible, 
but rather due to the fact that 
(a) changes in relative energies (bonding) are already near-linear
(by optimizing the reference geometry) with respect to alchemical coupling 
(effectively canceling higher order terms), 
and (b) inflection points can occur which lead to worse predictions for second order estimates.
For the interpolation of the pseudpotentials used in this study, 
inflection points near $\lambda=0$ are always observed 
when a lighter main group element is coupled to a heavier one. 
The absence of inflection points near $\lambda=1$ 
improves the predictive power of the second order correction:  
As such, the asymmetry of $\Delta E(d,d_0)$ with respect to $\lambda=0.5$
results in asymmetric predictive performance.

The choice of the reference geometry has a dramatic impact
on the predictive power of the alchemical estimates. 
For covalent bond potentials, a linear relationship has been identified, 
$(d_{\rm opt}\approx 0.76\:d_{\rm eq}^{\rm T} + 0.97\:{\rm \AA})$, 
that can be used to predict optimal $d_0$ requiring only rough estimates
of the equilibrium bond-length in the target molecule 
(which can easily be obtained using universal force-fields or semi-empirical
quantum chemistry methods).

We have found oscillating behaviour in the predictions of truncated Taylor series when 
varying the order in the non-vertical alchemical stretching of H$_2^+$. 
The crossing of eigenvalue surfaces is due to the electron density's necessity to be 
symmetric at $\lambda=0$ and $\lambda=1$.
This leads to a diverging series Taylor series, yet the second order correction could still provide fair predictions.
The behavior of first and second order truncated alchemical Taylor series expansions in non-vertical
transmutations in chemical space has also been analyzed for molecules with ten electrons.
Numerical evidence of the superior performance of $\Delta E^{(2)}_{\rm CP}$ over $\Delta E^{(2)}_{\rm IPA}$ 
suggests that the response of Coulomb and xc energy to alchemical perturbation is crucial. 

In summary, our findings indicate that a careful choice of alchemical interpolation paths 
enables alchemical derivatives to achieve predictive power with chemical accuracy for covalent bond potentials. 
Future work will deal with angles and torsions in lager molecules, as well as with
solid metals and ionic crystals.

\section{Acknowledgements}
We would like to thank K. Morokuma, E. Tapavicza, Q. Cui, A. Alavi, P. Ayers and P. Geerlings for discussions. 
OAvL acknowledges funding from the Swiss National Science foundation (No.~PP00P2\_138932). 
SF acknowledges the Research Foundation Flanders (FWO) for financial support.
Some calculations were performed at sciCORE (http://scicore.unibas.ch/) scientific computing core facility at University of Basel.

\bibliography{literature}{}
\bibliographystyle{ieeetr}

\end{document}